%% file: main.tex
\documentclass[sigconf]{acmart}

\usepackage{amsmath,amsfonts}
\usepackage[ruled,vlined]{algorithm2e}
\usepackage{tabularx}
\usepackage{subfig}
\usepackage{graphicx}
\usepackage{physics}
\usepackage{textcomp}
\usepackage{fancyhdr}
\usepackage{enumitem}
\usepackage{afterpage}
\usepackage{array}
\usepackage{bm}
\usepackage{float}
\usepackage[normalem]{ulem}
\usepackage{caption}
\usepackage{environ}
\usepackage{makecell}
\usepackage{textcomp}
\usepackage{booktabs}
\usepackage{lipsum}
\usepackage{pifont}
\newcolumntype{P}[1]{>{\centering\arraybackslash}p{#1}}
\colorlet{shadecolor}{gray!20}

\def\BibTeX{{\rm B\kern-.05em{\sc i\kern-.025em b}\kern-.08em
    T\kern-.1667em\lower.7ex\hbox{E}\kern-.125emX}}

\newcommand{\coo}{CO\textsubscript{2}}

\usepackage{tcolorbox,tikz}
\tcbuselibrary{skins}
\tcbuselibrary{breakable}

\newtcolorbox{mybox}[3][]
{
  breakable, 
  enhanced,
  colback  = #2!5, 
  colframe= #2!5,
  boxsep=-0.5mm,
  borderline west={1.5mm}{0.05mm}{#3!30}, 
  #1,
}

\newcounter{RQcounter}
\newcommand\showRQcounter{\stepcounter{RQcounter}\theRQcounter}

\newcounter{IQcounter}
\newcommand\showIQcounter{\stepcounter{IQcounter}\theIQcounter}

\renewcommand\footnotetextcopyrightpermission[1]{}

\begin{document}

\title{Toward Sustainable HPC: Carbon Footprint Estimation and Environmental Implications of HPC Systems}

\author{Baolin Li}
\affiliation{%
  \institution{Northeastern University}
  \country{}}
\author{Rohan Basu Roy}
\affiliation{%
  \institution{Northeastern University}
  \country{}}

\author{Daniel Wang}
\affiliation{%
  \institution{Northeastern University}
  \country{}}
  
\author{Siddharth Samsi}
\affiliation{%
  \institution{MIT}
  \country{}}

\author{Vijay Gadepally}
\affiliation{%
  \institution{MIT}
  \country{}}

\author{Devesh Tiwari}
\affiliation{%
  \institution{Northeastern University}
  \country{}}

\begin{abstract}
The rapid growth in demand for HPC systems has led to a rise in carbon footprint, which requires urgent intervention. In this work~\footnote{This work has been accepted at the 2023 ACM/IEEE International Conference for High Performance
Computing, Networking, Storage, and Analysis (SC '23)}, we present a comprehensive analysis of the carbon footprint of high-performance computing (HPC) systems, considering the carbon footprint during both the hardware manufacturing and system operational stages. Our work employs HPC hardware component carbon footprint modeling, regional carbon intensity analysis, and experimental characterization of the system life cycle to highlight the importance of quantifying the carbon footprint of HPC systems. 
\end{abstract}
\settopmatter{printacmref=false}


\maketitle
\pagestyle{empty}

\input{./sections/introduction.tex}

\input{./sections/background.tex}

\input{./sections/methodology.tex}

\input{./sections/part1.tex}
\input{./sections/part2.tex}

\input{./sections/part3.tex}

\input{./sections/related_work.tex}

\begin{acks}
This material is based upon work supported by the Assistant Secretary of Defense for Research and Engineering under Air Force Contract No. FA8702-15-D-0001, and United States Air Force Research Laboratory Cooperative Agreement Number FA8750-19-2-1000. Any opinions, findings, conclusions, or recommendations expressed in this material are those of the author(s) and do not necessarily reflect the views of the Assistant Secretary of Defense for Research and Engineering, or the United States Air Force. The U.S. Government is authorized to reproduce and distribute reprints for Government purposes notwithstanding any copyright notation herein. 
\end{acks}

\balance
\bibliographystyle{unsrtnat}
\bibliography{refs}

\end{document}

%% file: sections/introduction.tex
\section{Introduction}
\label{sec:intro}

High-performance computing (HPC) has become an essential tool in scientific research, engineering, and many other fields~\cite{portegies2020ecological,panda2021mvapich,feichtinger2011walberla}. The demand for HPC has experienced rapid growth in recent years. According to the US International Trade Commission, in 2010, there was reportedly 1.2 trillion gigabytes of new data created globally. However, this number is estimated to increase to a staggering 175 trillion gigabytes by 2025~\cite{usitc}. While the need for HPC resources is expanding, there is a downside to this growth. As more and more HPC systems are built and the size of these systems continues to increase~\cite{statista}, this leads to a rise in carbon footprint. For example, the Summit supercomputer built in 2017 has a peak power consumption of 13 MW, while in 2021, the next-generational Frontier supercomputer has more than doubled the peak power to 29MW~\cite{frontier_arch}. 

\vspace{3mm}

The carbon footprint of an HPC system cannot be accurately characterized by power consumption alone. The energy source used to power the system is also a key contributor to its environmental impact. Renewable energy sources such as hydropower and solar emit more than 20$\times$ less \coo{} than traditional energy sources like coal~\cite{gupta2022act}. Furthermore, a significant amount of carbon emission is incurred during the manufacturing and packaging of the HPC system components~\cite{chang2012totally,gupta2022chasing} before the system is deployed into operation. It is estimated that by 2030, datacenters and HPC systems may account for up to 8\% of the worldwide emissions if not intervened~\cite{andrae2015global}. As a result, major technology companies have been heavily invested in offsetting carbon emission in their datacenters~\cite{amazon_netzero,google_netzero,meta_netzero,apple_netzero}, while more and more research efforts have started to focus on the carbon-friendliness of large-scale systems~\cite{cao2022towards,acun2023carbon,gu2020energy, li2023green}. 

\vspace{3mm}
However, despite the current effort, there are still many unexplored and unanswered questions regarding HPC system sustainability and the carbon footprint of our HPC systems -- esp. from a procurement and operational point of view. One key question is how to quantify the carbon footprint of an HPC system holistically where the carbon emissions from the hardware manufacturing to the end of the system life cycle have all been accounted for. \textbf{In this work, we analyze the  carbon footprint from both the production and operational stages of an HPC system to address a series of unexplored investigations (referred to as Research Questions or RQs).} For each RQ, we conduct detailed modeling and characterization related to the question followed by visualization, discussion, and summarized takeaways. The current state of practice and available data, unfortunately, makes it very challenging to collect/build/analyze a standardized and portable model for carbon footprint accounting -- this paper aims to raise awareness and calls for a joint effort between vendors and HPC facilities to address this challenge. \textbf{The highlights of our analysis include the following:}

\vspace{3mm}

\textit{We perform modeling and analysis of the carbon emission during the production stage of individual HPC hardware components and compare their contributions.}  We highlight that performance benchmarking alone is not sufficient to achieve environmental sustainability. The contribution from the embodied carbon footprint of memory and storage devices in HPC systems cannot be ignored - while storage system has been traditionally viewed as a secondary optimization goal for performance and top 500 rankings, the carbon embodied in hard drives and solid storage drives presents a serious challenge to sustainability. As the HPC centers prepare to serve more memory-intensive scientific applications, they should carefully consider the hidden carbon cost of these memory modules. Carbon-conscious HPC facilities should explicitly request the embodied carbon specifications for all components from the chip vendor as a part of their request for proposal (RFP). 

\vspace{3mm}
\textit{Based on real-world carbon intensity data, we quantify the importance of regional carbon intensity when evaluating the operational carbon emissions of large-scale HPC systems.} We demonstrate, as expected, the average carbon intensity of energy sources varies across different regions, but surprisingly, even the regions with the lowest carbon intensity can have significant hourly temporal variations -- highlighting the significance of building cross-regional HPC systems. We identify a strong opportunity for systems researchers to design, develop, and deploy carbon-intensity-aware job schedulers to exploit temporal variations. Similar to core-hour accounting and budgeting, we recommend that HPC system operators and allocation programs should allocate a carbon budget to HPC users and some users could be prioritized to reduce their queue wait time if the carbon footprint of their jobs has been economical. 

\vspace{3mm}
\textit{From a holistic point of view, we integrate the two key aspects of sustainability: the modeling of carbon emission during production and the characterization of carbon emission during operation to analyze the environmental impact throughout the HPC system life cycle. } Our analysis reveals that hardware upgrades are often attractive from a performance point of view, but surprisingly, they can introduce significant embodied carbon that may not be offset quickly, especially if the center already runs on renewable energy sources or have low utilization. In the past, carbon-unaware system upgrades have not quantified and considered these factors. We provide a framework to help system practitioners make decisions on system upgrades based on hardware, workload, regional carbon intensity, performance, projected system lifetime, and user usage pattern.  

\vspace{3mm}

Our end-to-end modeling and characterization of an HPC system can serve as a stepping stone for future research in the carbon footprint perspective of HPC systems. HPC system practitioners can utilize our analysis to gain a better understanding of how sustainable the current system is and the layout of next-generational systems. To promote the idea of sustainable HPC and encourage more research efforts toward carbon neutrality in the community, our framework will be publically available. 

\vspace{2mm}

We acknowledge that this work is naturally prone to certain expected threats to validity because of the limited (but evolving) availability of accurate and widely-accepted carbon footprint data for various HPC components. As the community's awareness around carbon footprint increases, more robustness in modeling and characterization of carbon footprint will naturally evolve. Nevertheless, we hope that our study and analysis acts as the first step for raising carbon awareness in HPC practitioners and researchers to evaluate the carbon footprint of their systems, and help us grow as a community.

%% file: sections/background.tex
\section{Background and Methodology}
\label{sec:backg}


To analyze the carbon footprint of a system, we need to do so from two perspectives: the embodied carbon footprint and the operational carbon footprint. Embodied carbon refers to the carbon emissions associated with one-time expenditures like the production, transportation, and disposal of the materials and equipment used in HPC systems. In this work, we focus on modeling the production phase because the transportation and recycling of the component have been reported to be not dominant~\cite{gupta2022act} and tend to be consistent across different generations of the system. On the other hand, the embodied carbon's complementary is the operational carbon footprint, which refers to the day-to-day operation of a system. This includes the emissions associated with the electricity used to power the servers and other equipment, as well as the emissions associated with the cooling and ventilation systems used to keep the equipment within safe operating temperatures. In our work, we denote the overall carbon footprint, the embodied carbon footprint, and the operational carbon footprint as $C_{\text{total}}$, $C_{\text{em}}$, $C_{\text{op}}$, respectively. We can calculate the carbon footprint of a system using Eq.~\ref{eq:carbon_footprint}.
\begin{align}
    \label{eq:carbon_footprint}
    C_{\text{total}} = C_{\text{em}} + C_{\text{op}}
\end{align}

\subsection{Embodied Carbon Footprint Modeling}

The modeling of embodied carbon footprint is critical for the sustainability of semiconductor products~\cite{gupta2022act,chang2012totally,wang2008meeting}. We model the embodied carbon footprint of the HPC system components using principles similar to the ACT carbon modeling tool~\cite{gupta2022act} and inspired by early studies in the chip-level carbon modeling including Greenchip and ACT modeling efforts~\cite{jones2013considering,kline2016holistically,kline2019greenchip,gupta2022act}; some of the carbon modeling efforts related to manufacturing and fabrication were started as early as 2010s. We provide conceptual coverage of embodied carbon footprint modeling, but more details of the manufacturing carbon modeling are also available in prior studies~\cite{gupta2022act,kline2019greenchip}. Although these prior studies do not systematically model, characterize, derive insights, and identify challenges related to the carbon footprint of HPC systems.

Embodied carbon footprint is categorized into manufacturing carbon and packaging carbon. Manufacturing carbon refers to the emissions created from the creation of electronic components, such as transistors and resistors, from raw materials. Packaging carbon refers to the assembly of these components into functional chips and circuit boards. We summarize their relationships in Eq.~\ref{eq:embodied}.
\begin{align}
    \label{eq:embodied}
   C_{\text{em}} = \text{Manufacturing Carbon} + \text{Packaging Carbon} 
\end{align}

Modeling the manufacturing carbon footprint of different types of components requires a different approach. To quantify the manufacturing carbon of processors (i.e., CPUs, GPUs), we follow a vendor-generic approach to collect the part-specific information on die area ($A_{\text{die}}$), fab carbon emission per unit area ($FPA$, related to fab location and lithography), emissions from chemicals and gases per unit area ($GPA$, related to lithography), emissions from raw materials ($MPA$, related to lithography), and fab yield ($Yield$, set to a constant value of 0.875, consistent with~\cite{gupta2022act}) to estimate the amount of \coo{} emitted during the manufacturing process. We obtain such information from public product datasheets and sustainability reports. The manufacturing embodied carbon footprint of a processor unit ($M_{\text{proc}}$, unit: g\coo{}) can be calculated using Eq.~\ref{eq:embodied_pu}:
\begin{align}
    \label{eq:embodied_pu}
   M_{\text{proc}} = \frac{(FPA + GPA + MPA)\cdot A_{\text{die}}}{Yield}
\end{align}

We estimate the manufacturing carbon of memory and storage devices (DRAM, SSD, HDD) carefully in a vendor-specific way because these components have distinctive internal architectures. We first determine the capacity (e.g., GB) of a memory/storage device, and use publicly available sustainability reports of the vendor to estimate how much carbon is emitted per GB of the memory/storage device manufactured, denoted as emission per capacity ($EPC$). We can calculate the manufacturing footprint of a memory/storage device ($M_{\text{m/s}}$, unit: g\coo{}) as:
\begin{align}
    \label{eq:embodied_mem}
   M_{\text{m/s}} = EPC\cdot Capacity
\end{align}

We estimate the packaging carbon by counting the number of integrated circuit (IC) packages on the component and using an average packaging overhead of 150 g\coo{} per IC according to industry reports~\cite{spil,gupta2022act}. The packaging carbon (g\coo{}) is:
\begin{align}
    \label{eq:packaging}
   \text{Packaging Carbon} = 150\cdot\text{Number\_of\_ICs}
\end{align}

Note that Eq.~\ref{eq:packaging} is only applicable to our processor and memory components because this is non-trivial for storage components. To mitigate this issue, we compile data from industry reports on the packaging-to-manufacturing ratio from the vendor website~\cite{seagate}. 
\vspace{1.5mm}

\begin{table}
\small
\centering
\caption{Modeled individual components.}
\vspace{-4mm}
\scalebox{.85}{
\begin{tabular}{lllll}
\toprule
\textbf{Type} & \textbf{Component} & \textbf{Part Name} & \textbf{Release Date}\\ 
\midrule
\midrule
GPU & NVIDIA A100 & NVIDIA A100 PCIe 40GB & May 2020\\
\midrule
GPU & AMD MI250X & AND INSTINCT MI250X & November 2021\\
\midrule
GPU & NVIDIA V100 & NVIDIA V100 SXM2 32GB & March 2018\\
\midrule
CPU & AMD EPYC 7763 & AMD EPYC 7763 CPU & March 2021\\
\midrule
CPU & AMD EPYC 7742 & AMD EPYC 7742 CPU  & August 2019\\
\midrule
CPU & Intel Xeon Gold 6240R & Intel Xeon Gold 6240R CPU & February 2020\\
\midrule
DRAM & DRAM 64GB & SK Hynix 64GB DDR4 & October 2020\\
\midrule
SSD & SSD 3.2TB & Seagate Nytro 3530 3.2TB & October 2018\\
\midrule 
HDD & HDD 16TB & Seagate Exos x16 16TB & June 2019\\
\bottomrule
\end{tabular}}
\vspace{-4mm}
\label{table:components}
\end{table}

\noindent\textbf{Modeling of individual components. } We perform embodied carbon footprint modeling of individual components of an HPC system. We have listed the hardware components modeled in Table~\ref{table:components}. In the table, we have selected the frequently deployed GPU and CPU processors in the Top500 list~\cite{top500} from three major processor vendors: NVIDIA, AMD, and Intel. For example, the AMD MI250X GPU is available in the Frontier and LUMI supercomputers. We have chosen Sk Hynix and Seagate as the DRAM and storage vendors due to the availability of sustainability reports from these vendors~\cite{seagate,hynix}. Based on the vendor information, we have set the $EPC$ of DRAM, SSD, and HDD to 65 g\coo{}/GB, 6.21 g\coo{}/GB, and 1.33 g\coo{}/GB, respectively. 
\vspace{1.5mm}

\begin{table}
\small
\centering
\caption{Studied HPC Systems}
\vspace{-4mm}
\scalebox{.87}{
\begin{tabular}{lllll}
\toprule
\textbf{System} & \textbf{Location} & \textbf{CPU \& GPU} & \textbf{Cores} & \textbf{Year} \\ 
\midrule
\midrule
Frontier~\cite{frontier} & \makecell[l]{Oak Ridge, TN\\United States} & \makecell[l]{AMD EPYC 7763,\\AMD Instinct MI250X} & 8,730,112 & 2021 \\ 
\midrule
LUMI~\cite{LUMI} & \makecell[l]{Kajaani, Finland} & \makecell[l]{AMD EPYC 7763,\\AMD Instinct MI250X} & 2,220,288 & 2022 \\
\midrule
Perlmutter~\cite{perlmutter} & \makecell[l]{Berkeley, CA\\United States} & \makecell[l]{AMD EPYC 7763,\\NVIDIA A100 SXM4} & 761,856 & 2021\\ 
\bottomrule
\end{tabular}}
\vspace{-4mm}
\label{table:supercomputer}
\end{table}

\noindent\textbf{Modeling of HPC systems. } We conducted our study for state-of-the-art high-performance computing (HPC) systems to analyze the embodied carbon contribution of each component. Specifically, we analyzed the Frontier, LUMI, and Perlmutter supercomputers, as listed in Table~\ref{table:supercomputer}. We selected these systems because they are among the top 10 supercomputers in the Top500 list~\cite{top500} and were built in recent years. For each component present in the system, We calculate the $C_{\text{em}}$ according to Eq.~\ref{eq:embodied} and multiply it by the total number of components available.

\subsection{Operational Carbon Footprint Characterization}

The operational carbon footprint is characterized when workloads are running on the system. It can be calculated using the carbon intensity of the power plant that powers the system ($I_{\text{sys}}$, unit: g\coo{}/kWh) and the system's operational energy ($E_{\text{op}}$, unit: kWh). 
\begin{align}
    \label{eq:operational}
   C_\text{op} = I_\text{sys}\cdot E_\text{op}
\end{align}

Carbon intensity $I_{\text{sys}}$ is a metric of how many grams of $CO_2$ are released into the atmosphere to produce a unit of energy. It depends on the fuel mix from the power plant. Higher carbon intensity means that the energy source generates more carbon emissions when producing the same amount of energy. Sustainable sources of energy such as wind or solar have a carbon intensity of less than 50 g\coo{}/kWh while non-renewable sources like coal have a carbon intensity of more than 800 g\coo{}/kWh. The operational energy ($C_\text{op}$) is the product of the IC component energy and the HPC system power-usage-effectiveness (PUE), which we set to a constant across all systems we characterize. In this work, we use the carbontracker~\cite{anthony2020carbontracker} tool to measure a system's operational carbon footprint $C_{\text{op}}$ while running certain benchmark suites.

%% file: sections/methodology.tex
\begin{table}
\small
\setlength\defaultaddspace{0.66ex}
\centering
\caption{Independent system operators and regions.}
\vspace{-4mm}
\scalebox{0.9}{
\begin{tabular}{lll}
\toprule
\makecell[l]{\textbf{Operator}\\\textbf{Name}} & \makecell[l]{\textbf{Country of}\\\textbf{Operation}} & \makecell[l]{\textbf{Region of}\\\textbf{Operation}} \\ 
\midrule
\midrule
Kansai (\textbf{KN})~\cite{kn} & Japan & Kansai Region \\
\midrule
Tokyo (\textbf{TK})~\cite{tk} & Japan & Tokyo Region \\
\midrule
\makecell[l]{Electricity System Operator (\textbf{ESO})~\cite{eso}} & United Kingdom & Great Britain\\
\midrule
\makecell[l]{California Independent\\System Operator (\textbf{CISO})~\cite{ciso}} & United States & California\\
\midrule
\makecell[l]{Pennsylvania-New Jersey-Maryland\\Interconnection (\textbf{PJM})~\cite{pjm}} & United States & Mid-Atlantic US \\
\midrule
 \makecell[l]{Midcontinent Independent\\System Operator (\textbf{MISO})~\cite{miso}} & \makecell[l]{United States,\\Canada} & \makecell[l]{Midwest US,\\Manitoba}\\
\midrule
\makecell[l]{Electric Reliability\\Council of Texas (\textbf{ERCOT})~\cite{ercot}} & United States & Texas \\
\bottomrule
\end{tabular}}
\vspace{-4mm}
\label{table:operators}
\end{table}

\vspace{1.5mm}
\noindent\textbf{Geographical carbon intensity. } According to Eq.~\ref{eq:operational}, the operational carbon footprint is proportional to the carbon intensity which highly depends on the time and the geographical location. In our work, we study the carbon intensity across different geographical regions to get a better understanding of the operational carbon footprint of a system. We have collected carbon intensity data from multiple power system operators across the globe as listed in Table~\ref{table:operators}. We obtain the ESO (UK) data from ESO's public Carbon Intensity API~\cite{eso_ci} and other regions' carbon intensity from Electricity Maps~\cite{electricity_map}. For all the regions in Table~\ref{table:operators}, we perform carbon intensity analysis on hourly data (year 2021).

\begin{table}
\centering
\caption{Benchmarks performed and their respective models.}
\vspace{-4mm}
\scalebox{0.9}{
\begin{tabular}{ll}
\toprule
\textbf{Benchmark} & \textbf{Models}\\ 
\midrule
\midrule
\makecell[l]{Natural Language\\Processing (NLP)} & \makecell[l]{BERT~\cite{devlin2018bert}, DistilBERT~\cite{sanh2019distilbert},\\MPNet~\cite{song2020mpnet}, RoBERTa~\cite{liu2019roberta}, BART~\cite{lewis2019bart}} \\
\midrule
Computer Vision (Vision) & \makecell[l]{ResNet50~\cite{he2016deep}, ResNext50~\cite{xie2017aggregated},\\ShuffleNetV2~\cite{ma2018shufflenet}, VGG19~\cite{simonyan2014very}, ViT~\cite{mao2022towards}} \\
\midrule
CANDLE~\cite{Candle2022,wu2019performance} & Combo, NT3, P1B1, ST1, TC1 \\
\bottomrule
\end{tabular}}
\vspace{-4mm}
\label{table:benchmarks}
\end{table}

\vspace{1.5mm}
\noindent\textbf{Benchmarking workloads. } To calculate the operational carbon, we also need to characterize the operational energy when the system is running. In this work, we perform a benchmarking study on real systems of different generations. By benchmarking on representative workloads, we are able to compare the operational carbon footprint of different systems and their respective performance. We have listed the details of the benchmark sets in Table~\ref{table:benchmarks}. These benchmark sets represent the deep learning training workload across different research fields. We choose deep learning training because this is the target workload of today's GPU systems. The NLP benchmarks are provided by Huggingface, where we perform the question-answering task on various language models. The Vision benchmarks are provided by Pytorch, and we select models that have highly varied architecture (e.g., residual network in ResNet50, transformer in ViT) to perform image classification. The CANDLE benchmarks are provided by Argonne National Laboratory (ANL), with the initiative to address cancer research challenges with deep learning. We select five benchmarks from the Pilot1 class which represent problems in predicting drug response based on molecular features of tumor cells and drug descriptors.

Next, we provide a holistic carbon footprint analysis from three perspectives of HPC systems: (i) embodied carbon footprint, (ii) geographical carbon intensity variation, and (iii) total carbon footprint and performance benchmarking of real-world workloads.

%% file: sections/part1.tex
\section{Embodied Carbon Analysis }
\label{sec:p1}

First, we analyze the relative embodied carbon footprint for different HPC system components. We have listed the details of our modeled components in Table~\ref{table:components}. These components appear frequently on the top 500 supercomputer list, represent a wide diversity in terms of vendors and time, and have their carbon-related specifications accessible or derivable. 

\begin{mybox}{blue}{blue}
\textbf{RQ \showRQcounter.} \textit{How does embodied carbon vary among different types of GPUs, CPUs, and memory/storage? How does the embodied carbon vary after being normalized to performance?}
\end{mybox}

\begin{figure}[t]
    \centering
    \includegraphics[scale=0.36]{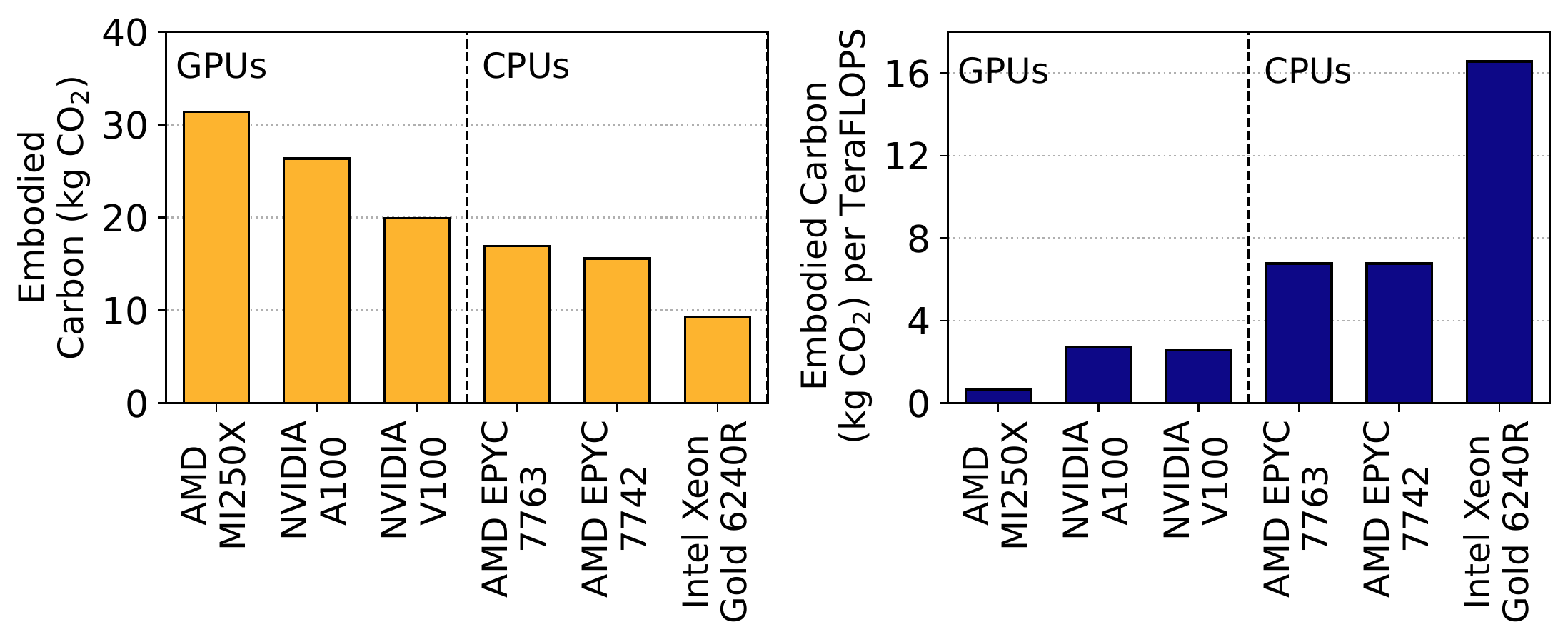}
    \vspace{-4mm}
    \caption{Embodied carbon footprint of GPU/CPU devices, and the footprint normalized to theoretical double-precision floating point performance (TeraFLOPS).}
    \label{fig:cpugpu_embidied_carbon}
    \vspace{-5mm}
\end{figure}

\begin{figure}[t]
    \centering
    \includegraphics[scale=0.36]{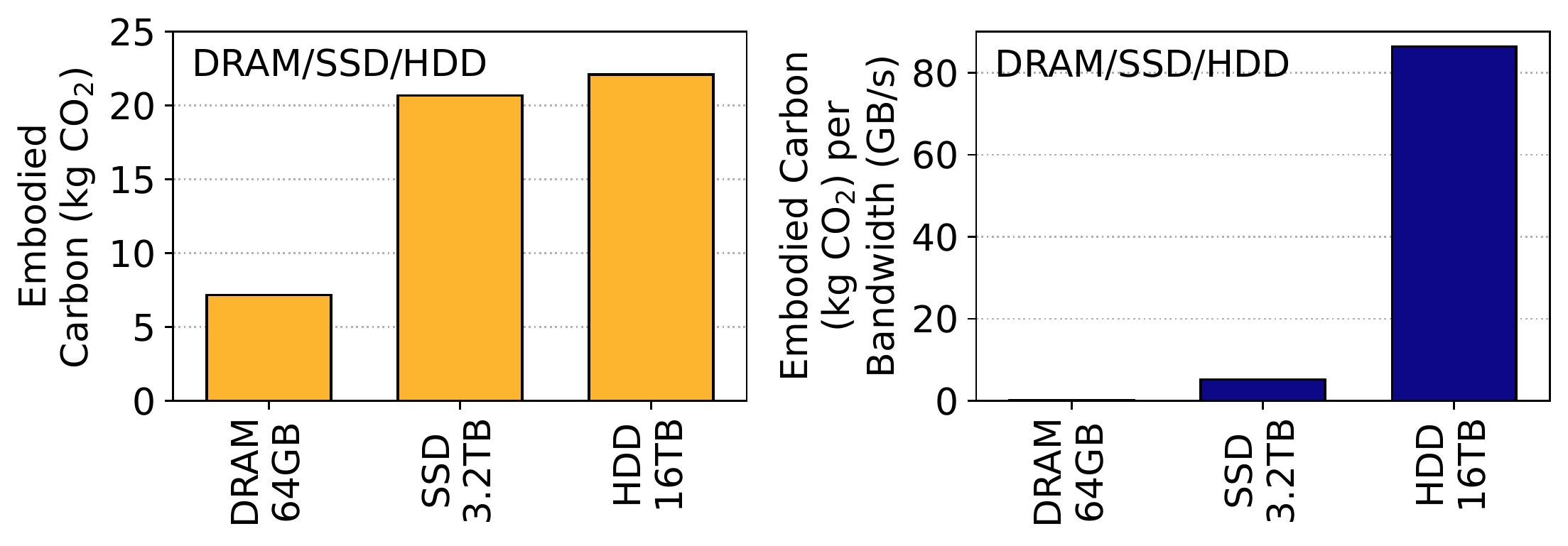}
    \vspace{-4mm}
    \caption{Embodied carbon footprint of DRAM/SSD/HDD devices and the footprint normalized to bandwidth (GB/s).}
    \label{fig:memhdd_embidied_carbon}
    \vspace{-5mm}
\end{figure}

\noindent\textbf{Result and Analysis.} In Fig.~\ref{fig:cpugpu_embidied_carbon}, we compare the embodied carbon footprint of the GPU components and CPU components we study in Table~\ref{table:components}. Fig.~\ref{fig:cpugpu_embidied_carbon} (a) shows that each GPU devices have higher embodied carbon than the CPU devices by up to 3.4$\times$. On the other hand, when we normalize the embodied carbon to the FP64 operation performance in Fig.~\ref{fig:cpugpu_embidied_carbon} (b), the trend is reversed: each CPU device has higher embodied carbon per FLOPS than any of the GPU devices. This is because CPUs provide much lower performance despite containing less embodied carbon and hence, they are not able to offset the lower performance with their embodied carbon efficiency. We have also observed similar trends on other floating point precisions such as the FP32.

The AMD MI250X GPU has the highest embodied carbon but has the lowest embodied carbon per FLOPS among all devices. This is because AMD has reported this GPU to have almost 5$\times$ higher peak FP64 FLOPS than an NVIDIA A100~\cite{amd-mi250x}. Note that we used FLOPS simply because it is the most commonly documented and used. However, our methodology is not specific or limited to FLOPS only; other figures of merit can be used for normalization as well.

\begin{mybox}{green}{green}
\textbf{Observation \showIQcounter.} We observe that GPUs tend to have significantly more embodied carbon than CPUs -- this seems to be true across multiple types of CPUs and GPUs that are used in different top 500 supercomputers. Although GPUs tend to have higher overall embodied carbon, the embodied carbon normalized to raw performance (g\coo{}/FLOPS) is lower than CPUs. 
\end{mybox}

Next, we investigate the embodied carbon of memory and storage devices in a system, as presented in Table~\ref{table:components}. From Fig.~\ref{fig:memhdd_embidied_carbon} (a), we can observe that each DRAM/SSD/HDD device has an embodied carbon of 5 to 25 kg\coo{}, which is in a comparable range to the GPU/CPU devices. Similar to the normalized analysis on GPUs/CPUs, we are also interested in the embodied carbon per bandwidth (GB/s), which is considered a key metric of memory/storage devices. The trend we observe in Fig.~\ref{fig:memhdd_embidied_carbon} (b) indicates that the embodied carbon per bandwidth of DRAM devices is significantly smaller than SSDs, and is negligible compared to the HDD devices due to a much higher DRAM bandwidth.

\begin{mybox}{green}{green}
\textbf{Observation \showIQcounter.} A typical single unit of memory and storage device also tends to have a comparable amount of embodied carbon as compute units (CPU/GPU), but it should be kept in mind that the capacity of memory and storage devices can affect the embodied carbon. 
\end{mybox}

\begin{mybox}{yellow}{yellow}
\textbf{Implication.} The implication is that carbon-conscious HPC facilities should explicitly request the embodied carbon specifications for CPUs and other computer accelerators from the chip vendor as a part of their request for proposal (RFP), in addition to performance benchmarking numbers. Performance benchmarking alone is not sufficient to achieve environmental sustainability. The embodied carbon footprint of memory and storage devices cannot be ignored either. While storage system has been traditionally viewed as a secondary optimization goal for performance and top 500 rankings, the carbon embodied in hard drives and solid storage drives can present a serious challenge to sustainability. 
\end{mybox}

\begin{figure}[t]
    \centering
    \includegraphics[scale=0.335]{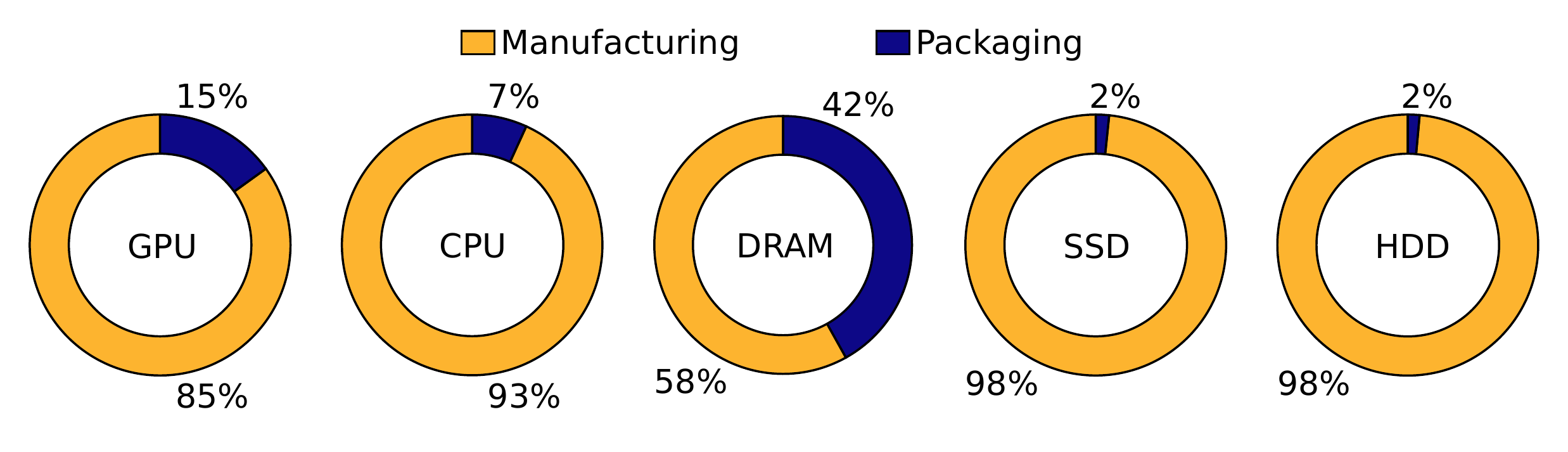}
    \vspace{-9mm}
    \caption{The manufacturing and packaging carbon footprint of the embodied carbon varies with device types.}
    \label{fig:pie_embidied_carbon}
    \vspace{-5mm}
\end{figure}

\begin{mybox}{blue}{blue}
\textbf{RQ \showRQcounter.} \textit{What is the breakdown of the embodied carbon for different types of GPUs, CPUs, and memory/storage, in terms of manufacturing and packaging carbon? }
\end{mybox}

\noindent\textbf{Result and Analysis. } In Fig.~\ref{fig:pie_embidied_carbon}, we quantify the carbon footprint associated with the manufacturing and packaging of various computer hardware components. The manufacturing footprint is incurred during the semiconductor wafer fabrication, assembly, and testing, while the packaging footprint represents the emission during the chip packaging process. 

Specifically, we focus on the carbon footprint of GPU, CPU, DRAM, SSD, and HDD. Our analysis is presented through a series of ring charts, where each chart represents the carbon footprint of one specific component. The charts are divided into two parts, representing the manufacturing and packaging carbon footprint. 

Interestingly, we found that the composition of the carbon footprint varied significantly between different components. For example, while the embodied carbon footprint of SSD and HDD was dominated by the manufacturing process, the packaging carbon footprint of DRAM was found to be 42\% of its overall embodied carbon footprint. This is due to the fact that DRAM chips are typically smaller and require more precise and delicate packaging due to their sensitivity to external factors such as temperature, humidity, and electrostatic discharge whereas SSDs and HDDs are larger and require less intricate packaging. The GPUs and CPUs have about 10\% of their embodied carbon from packaging, less than DRAM as potentially their manufacturing is more complex from factors such as lower lithography and larger die areas. 

\begin{mybox}{green}{green}
\textbf{Observation \showIQcounter.} As expected, the manufacturing carbon is the most dominant part of the embodied carbon for most components including GPUs, CPUs, HDDs, and SSDs. However, for DRAM, packaging carbon contributes over 40\% of the embodied carbon -- contributing toward a higher packaging-induced embodied carbon. 
\end{mybox}


\noindent\textbf{Limitation of this study.} We recognize that a complex HPC system tends to have additional components. Specifically, network interconnects such as HPE Slingshot~\cite{slingshot} provide high-bandwidth, low-latency communication between nodes; in a distributed file system, storage devices are connected to storage servers that are in turn connected to compute nodes~\cite{braam2019lustre}. In this work, these components could not be modeled and characterized due to the unavailability of open-access production carbon emission reports.  

\begin{mybox}{yellow}{yellow}
\textbf{Implication.} There is a critical shortage of carbon footprint data related to different components including networking equipments in HPC systems. HPC practitioners and vendors should work together to build standardized models for collecting and sharing embodied carbon of different components. 
\end{mybox}

\begin{mybox}{blue}{blue}
\textbf{RQ \showRQcounter.} \textit{How do the embodied carbon and performance vary for different workloads as the number of GPUs increases in a compute node?}
\end{mybox}

\begin{figure}[t]
    \centering
    \includegraphics[scale=0.36]{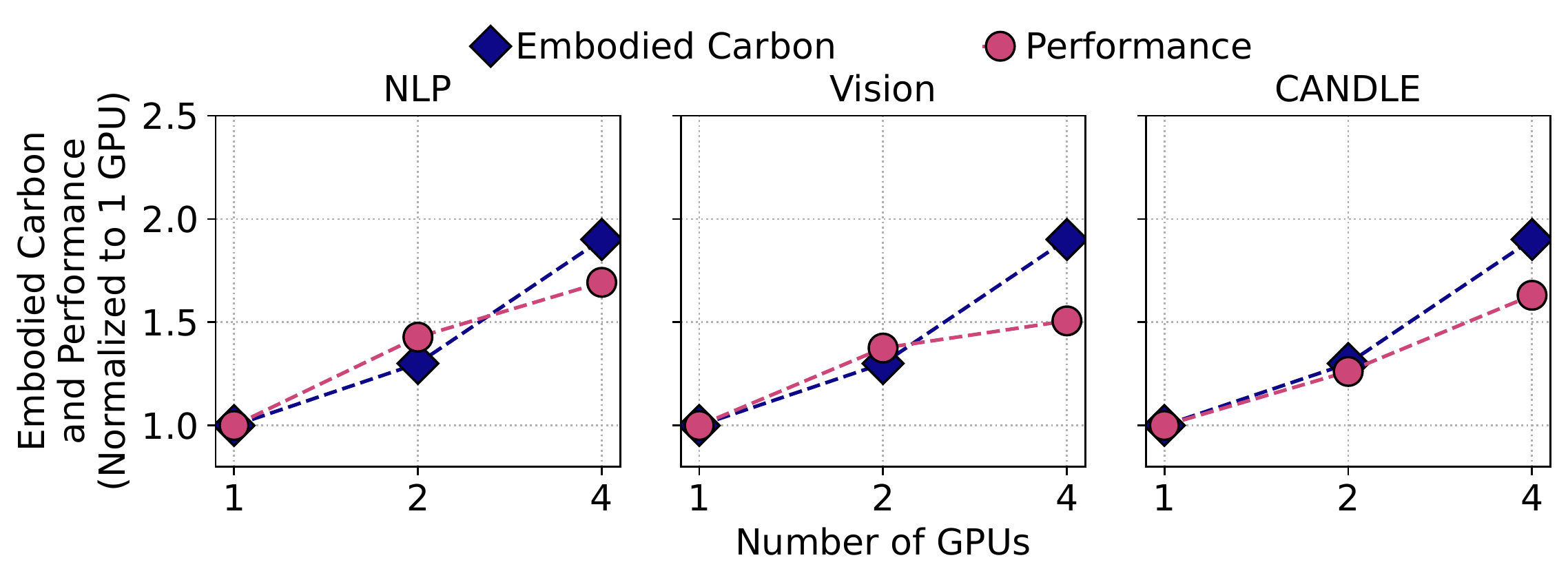}
    \vspace{-5mm}
    \caption{Increasing the number of GPUs in a node can improve node performance, but it also leads to a higher embodied carbon footprint. However, as expected, the performance gains tend to plateau, but the carbon footprint continues to increase.}
    \vspace{-5mm}
    \label{fig:cpu_gpu_ratio}
\end{figure}

\noindent\textbf{Result and Analysis.} In Fig~\ref{fig:cpu_gpu_ratio}, show how the embodied carbon footprint and the system performance vary when we increase the number of GPUs in a node. In our node, we have two Intel Xeon Gold 6240R CPUs, and we vary the number of NVIDIA V100 GPUs between 1, 2, and 4. We characterize the 1-GPU, 2-GPU, and 4-GPU system performance using the three sets of benchmarks in Sec.~\ref{sec:backg} (Table~\ref{table:benchmarks}). We compare the performance against the embodied carbon footprint of the node. We have kept the batch size per GPU in these benchmarks consistent as we increase the number of GPUs. 

As expected, the embodied carbon footprint increase is proportional to the number of GPUs added. For all benchmark sets, when we increase the number of GPUs to 2, both the embodied carbon and the node performance are increased by approximately 30\% to 40\% of the normalized carbon footprint and the corresponding performance, meaning the performance-to-embodied-carbon ratio is approximately 1. However, as we further increase the number of GPUs to 4, the performance increase cannot keep up with the embodied carbon footprint due to heavier communication overhead between the GPUs, and the performance-to-embodied-carbon ratio has dropped to approximately 0.88 for the NLP and CANDLE benchmarks, and 0.79 for the Vision benchmarks.

\begin{mybox}{green}{green}
\textbf{Observation \showIQcounter.} Our experimental results demonstrate that the carbon footprint per unit of achieved performance of a workload may get worse as we increase the number of GPUs.   As expected, with increasing the number of GPUs, the performance may not increase linearly, but the embodied carbon increases linearly. Therefore, the overall carbon footprint per unit of achieved performance may increase. 
\end{mybox}

\begin{figure}[t]
    \centering
    \includegraphics[scale=0.35]{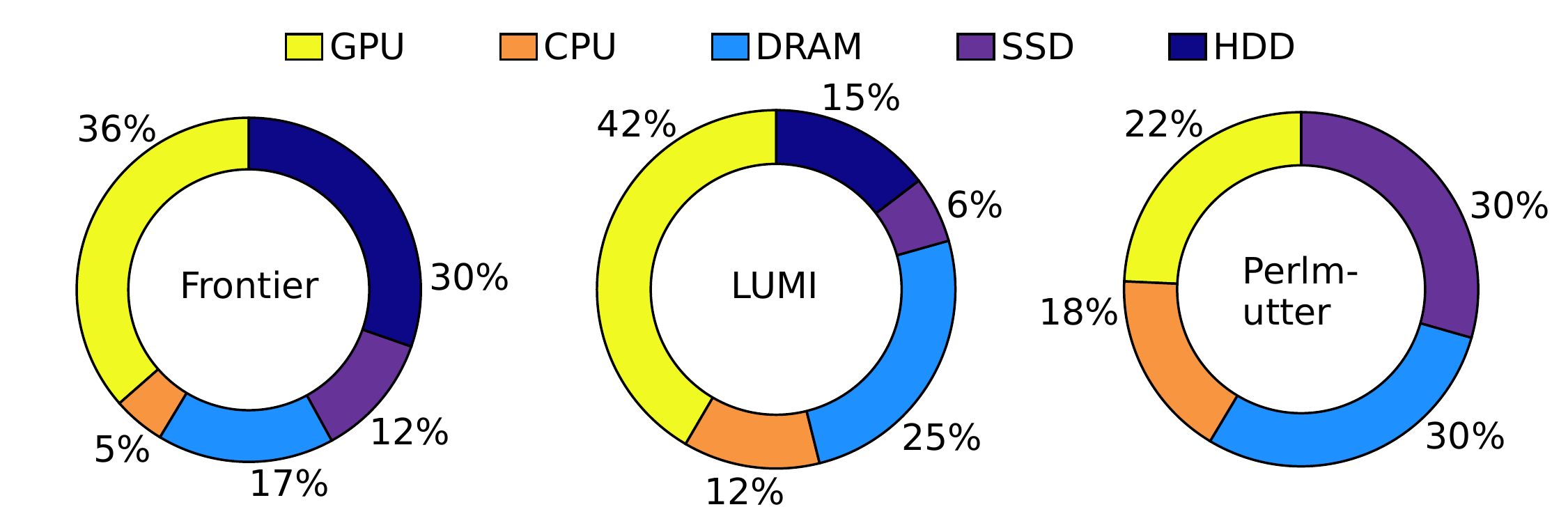}
    \vspace{-5mm}
    \caption{Carbon footprint contribution from different components in three leadership supercomputers: Frontier, LUMI, and Perlmutter.}
    \label{fig:hpc_carbon}
    \vspace{-5mm}
\end{figure}

\begin{mybox}{blue}{blue}
\textbf{RQ \showRQcounter.} \textit{For leading supercomputers, does the contribution from different components toward the overall embodied carbon change? Which is the most dominant embodied carbon in today's supercomputers: GPU, CPU, memory, SSD, or HDD?}
\end{mybox}

\noindent\textbf{Result and Analysis. } To better understand the sources and distribution of embodied carbon in supercomputing, we present a comparative analysis of three of the world's powerful supercomputers: Frontier, LUMI, and Perlmutter that are ranked $1^{st}$, $3^{rd}$, and $8^{th}$ respectively in the latest Top-500 list as of November 2022~\cite{top500}. We have listed a more detailed description of these systems in Table~\ref{table:supercomputer}. 

We note that the magnitude of the absolute carbon footprint of each supercomputer is not listed because it is not our intent to showcase that one is better than the other, or even compare them. Instead, we want to highlight that the composition of a system greatly affects the embodied carbon footprint breakdown. 

In Fig.~\ref{fig:hpc_carbon}, each ring chart shows the proportion of carbon footprint contributed by different components of the system, including the CPU, GPU, DRAM, SSD, and HDD. By examining the relative contributions of these components, we can gain insights into the amount of carbon emission embodied in the components when building different supercomputer architectures. We recognize that even the current systems can potentially undergo upgrades, the carbon footprint estimates of their components are subject to inaccuracy and limited to publicly available information, and these can change over time/generations. Therefore, instead of absolute numbers, we are more interested in projected estimates about how system composition can shift the carbon bottlenecks - this also guided our choice of selected supercomputers.

First, our analysis reveals interesting carbon footprint differences between the three selected supercomputers in terms of their composition. For example, while Frontier and LUMI have significantly more sizeable embodied carbon footprints due to their GPUs, Perlmutter has a more balanced embodied carbon distribution between CPUs and GPUs. This is because Perlmutter has a large CPU partition whereas LUMI has a relatively small CPU partition and Frontier is also relatively GPU-heavy. 

Second, the proportion of carbon emissions from storage devices varies between the three systems, reflecting differences in their storage architectures. Frontier has 695 PB of HDD storage that makes up large embodied carbon footprint, while Perlmutter deploys an all-flash file system. This is particularly important and highlights that the carbon contributions from storage can be significant and should be treated as a first-class citizen, even though they often take the backseat in performance rankings.

Interestingly, when we compare the compute components (CPU and GPU) against the memory and storage devices (DRAM, SSD, HDD) in terms of their embodied carbon, the memory and storage have made up approximately 60\% of the carbon in Frontier and Perlmutter, and almost 50\% in LUMI. This indicates that although these memory and storage devices do not consume as much power as compute devices, they inherently have a higher embodied carbon footprint, which cannot be neglected when building a supercomputer. This is in alignment with multiple concurrent and recent studies which have highlighted the importance of the carbon footprint of storage devices~\cite{tannu2022dirty,zuck2023degrading,mersy2023toward}. 

Finally, in Fig.~\ref{fig:hpc_carbon}, the GPUs have consistently higher embodied carbon footprint than CPUs in all three supercomputers, especially in Frontier, where the embodied carbon in GPUs is more than 7$\times$ that of the CPUs. This shows a trend that the GPUs are becoming important in today's supercomputers, not just from the performance point-of-view, but also for their carbon footprint.

\begin{mybox}{green}{green}
\textbf{Observation \showIQcounter.} Our analysis reveals that the breakdown of their embodied carbon differs significantly among different supercomputers -- even though, these supercomputers appear to be comparable in peak raw performance (among the top ten in the top 500 supercomputer list). Depending upon the supercomputer architecture and organization, GPU, memory or even SSD can be the most dominating factor. Surprisingly, DRAM contributes significantly to overall embodied carbon for all evaluated supercomputers. This is in contrast with an earlier result, where we showed that a DRAM card's embodied carbon was lower compared to a single CPU/GPU. The reason for an overall relatively higher contribution of DRAM is because of the high number of DRAM cards in the system -- which adds up.
\end{mybox}

\begin{mybox}{yellow}{yellow}
\textbf{Implication.} The implication is that HPC facilities should explicitly document and understand which components are contributing to their overall embodied carbon footprint. Currently, there is limited awareness of these factors in supercomputing center design. As energy sources powering the supercomputers become ``greener'', this aspect will become the most dominant factor in the overall carbon footprint of a supercomputing center. The memory footprint of computational science applications has been on rapid rise~\cite{rajbhandari2020zero,choukse2020buddy,peng2020demystifying}. Many supercomputing centers provision a large among of memory to serve such applications. As the HPC centers prepare to serve more memory-intensive scientific applications, they should carefully consider the hidden carbon cost of these memory modules. Memory often has the largest failure rate and gets replaced~\cite{ren2020exploring,zhang2019quantifying}, therefore, lack of attention around minimizing or mitigating embodied carbon cost for DRAM can be undesirable.
\end{mybox}



%% file: sections/part2.tex
\section{Geographical Carbon Intensity}
\label{sec:p2}

In Sec.~\ref{sec:p1}, we delved into the analysis of embodied carbon, which refers to the carbon emissions associated with the manufacturing and packaging of individual devices. Now, in this section, we shift our focus to operational carbon, which relates to the emissions resulting from the day-to-day operation of a large-scale system. It is worth noting that operational carbon strongly depends on regional carbon intensity, which reflects the amount of carbon emissions associated with electricity production in a given region. This means that datacenters and supercomputers located in regions with high carbon intensity will have a higher operational carbon footprint than those located in regions with lower carbon intensity. Therefore, it's crucial to analyze the regional carbon intensity when evaluating the operational carbon emissions of a system.

\begin{figure}[t]
    \centering
    \includegraphics[scale=0.36]{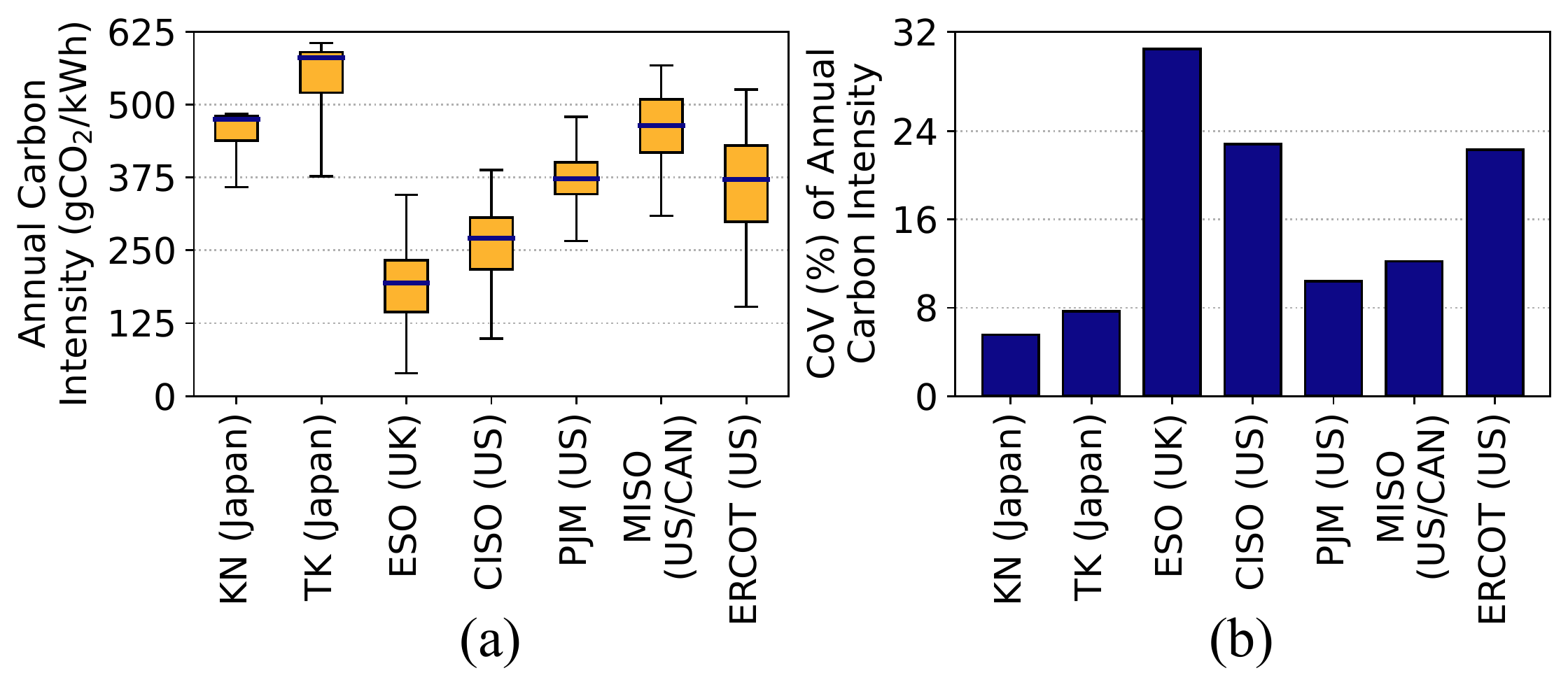}
    \vspace{-4mm}
    \caption{The average carbon intensity of the energy sources differs across geographical regions in Table~\ref{table:operators}, and a significant temporal variation exists for each geographical region. }
    \vspace{-5mm}
    \label{fig:regionwise_carbon_intensity}
\end{figure}

\begin{mybox}{blue}{blue}
\textbf{RQ \showRQcounter.} \textit{How does the carbon intensity vary across geographical regions?}
\end{mybox}

\noindent\textbf{Result and Analysis.} In Fig.~\ref{fig:regionwise_carbon_intensity}, we illustrate the annual carbon intensity in the year 2021 of seven different power system operators distributed across different countries and regions. We have provided more details of the system operators in Table~\ref{table:operators}. In Fig.~\ref{fig:regionwise_carbon_intensity} (a), we use a box plot to compare the annual carbon intensity for different regions, which display distinctive carbon intensity trends. 

Overall, the ESO (Great Britain, UK) region has the lowest carbon intensity among all regions, with a median carbon intensity of less than 200 g\coo{}/kWh. The TK (Tokyo, Japan) region has the highest carbon intensity among all regions, whose medium annual carbon intensity is three times ESO's. In recent years, there has been a growing interest in assessing the environmental impact of supercomputers, for example, the Green500 list~\cite{green500}. However, it is important to note that the carbon intensity of electricity generation can vary significantly among different regions. Therefore, when comparing the "greenness" of supercomputers, it is crucial to take into account the geographical location of the facility. This is because the electricity grid mix in one region may be heavily reliant on fossil fuels, while in another region, it may be predominantly powered by renewable energy sources. As such, a supercomputer located in a region with a higher proportion of renewable energy in its electricity grid mix would have a lower operational carbon footprint than a supercomputer located in a region with a higher proportion of fossil fuel-based electricity. Thus, in order to accurately evaluate the environmental impact of supercomputers, it is important to consider the carbon intensity difference among different regions.

Remarkably, Fig.~\ref{fig:regionwise_carbon_intensity} (b) shows a different side of the story when we show the coefficient of variation (CoV) in \%, which represents the standard deviation as a percentage of the average carbon intensity in the region. The two regions with the lowest medium carbon intensity -- ESO (Great Britain, UK) and CISO (California, US), also have the most variations in their carbon intensity. On the other hand, the regions with the highest medium carbon intensity -- TK (Tokyo, Japan) and KN (Kansai, Japan) have the least carbon intensity variation among all regions. This means that while the average carbon intensity of a region may be relatively low, there may be significant fluctuations in carbon emissions over time. One factor that contributes to variation in carbon intensity is the use of intermittent renewable energy sources such as wind or solar power, which may result in fluctuations in carbon emissions.

\begin{mybox}{green}{green}
\textbf{Insight \showIQcounter.} As expected, the average carbon intensity of the energy sources differs across geographical regions. On average,  the ESO (Great Britain, UK) region overall and the California region within the USA has the lowest carbon intensity. But, interestingly, the temporal variance in those regions is among the highest. Therefore, simply building a data center in the least-carbon intensity region is not an optimal solution at all times -- due to significant temporal variation. 
\end{mybox}

\begin{mybox}{yellow}{yellow}
\textbf{Implication.} The implication is that the federal agencies within a country and across countries should continue to strongly consider deploying similar architecture supercomputers across multiple geographical regions. Incidentally, this model has been followed in the past in the USA for other reasons (programmability, staffing, science missions). Our analysis reinforces that model from a carbon-aware computing perspective -- esp. in the collaborative international context, too. When ranking supercomputers based on their "greenness" (Green 500 ranking), we should also consider the geographical location of the facility and energy-mix, and its temporal variations -- which is not currently practiced. 
\end{mybox}

\begin{figure}[t]
    \centering
    \includegraphics[scale=0.36]{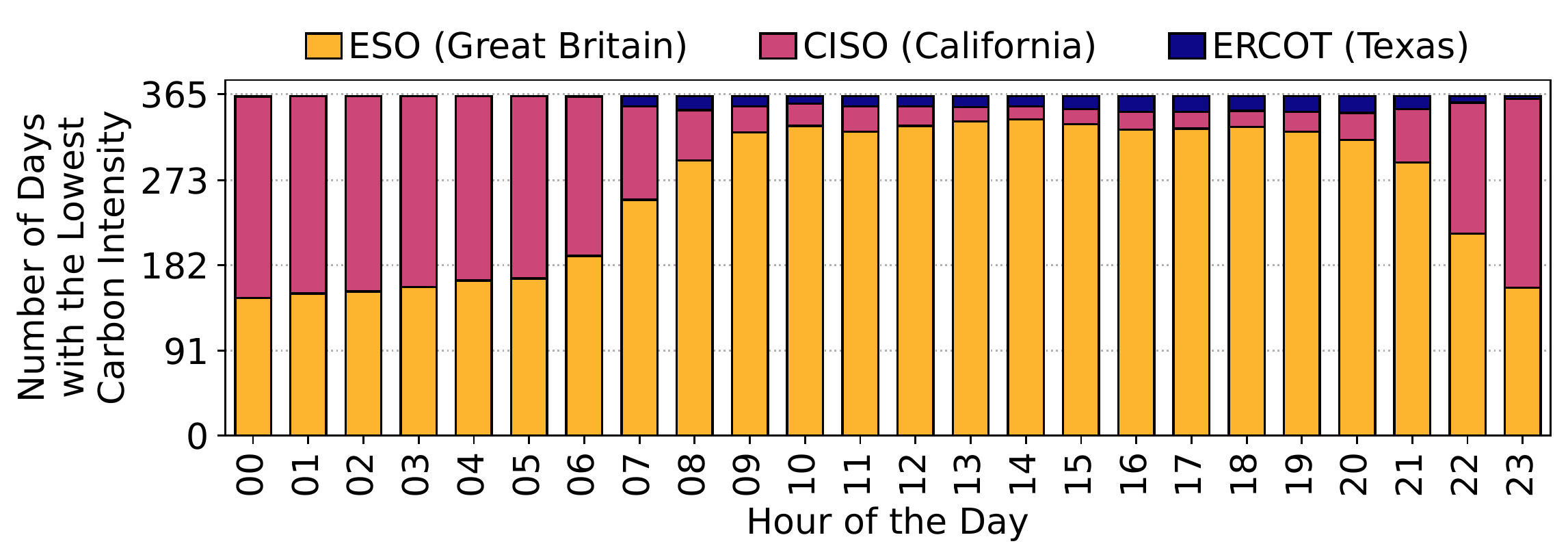}
    \vspace{-0.5cm}
    \caption{Hourly variation in carbon intensity across three most carbon-friendly regions. Although not shown explicitly, season variations also naturally exist.}
    \label{fig:hour_of_day_lowest_carbon}
    \vspace{-5mm}
\end{figure}

\begin{mybox}{blue}{blue}
\textbf{RQ \showRQcounter.} \textit{Can the temporal variation in carbon intensity be exploited at finer timescale (e.g., hours)?}
\end{mybox}

\noindent\textbf{Result and Analysis.} In Fig.~\ref{fig:hour_of_day_lowest_carbon}, we pick the three operator regions with the lowest medium carbon intensity and compare their carbon intensities during the same hour of the day. Since they are distributed in different geographical locations, we account for the difference between time zones (GMT, PST, CST) and convert them to JST (UTC+9) time during the analysis. On the y-axis, we show the number of days that the region experiences the lowest carbon intensity among all regions during that hour. For example, for the $1^{st}$ hour of the day, ESO (Great Britain) is the region with the lowest carbon intensity in about 150 days out of a year while CISO (California) is the lowest carbon intensity region in about 215 days out of a year. 

Fig.~\ref{fig:hour_of_day_lowest_carbon} reveals that the number of days that each region has the lowest carbon intensity during a given hour varies significantly throughout the year, with no region consistently having the lowest carbon intensity for any given hour. This is due to the different energy generation mixes and demands in different time zones in different regions. The hours during which ESO (Great Britain) is the region with the lowest carbon intensity, hour 8 to hour 20, are the midnight to noon time in the UK when electricity demand is expected to be low. Therefore, it would be beneficial to have more jobs running in the HPC centers in the ESO to exploit the availability of renewable energy. However, during other hours, running the job in the ESO region is more likely to yield more carbon emissions as CISO (California) is a ``greener'' region during most of the days. Overall, our analysis highlights the temporal variability in the carbon intensity of different regions and underscores the need for flexible and dynamic approaches to exploit the opportunity in distributing jobs across regions.

While visually not depicted, we verified that even when two regions have very similar carbon intensity (e.g. Mid-Atlantic US and Texas), it is possible to optimize for carbon footprint further by distributing jobs between data centers in these regions. This is because the regions exhibit temporal variations and these variations are aligned due to geographical characteristics -- for example, when wind power is more easily available in Texas, it may be not available in New Jersey at the same time.

\begin{mybox}{green}{green}
\textbf{Insight \showIQcounter.} Our analysis reveals that even among the greenest region, there is a significant hourly variation in carbon intensity. This variation is strong enough that no single region is a consistent winner for all hours of the day for all days in a year, and the number of days they are a winner in a year also varies -- motivating a case for geographically distributed data centers where jobs can be distributed. However, exploiting this opportunity is not trivial since the temporal variation on different days of the year varies. There are additional challenges related to latency and the energy consumption associated with data transfers when distributing the workload across geographically distributed HPC centers. Therefore, workload distribution policies should consider such a tradeoff.

\end{mybox}

\begin{mybox}{yellow}{yellow}
\textbf{Implication.} An important implication around these observations is an incentive structure for end users to exploit these fine-grained carbon intensity patterns. In particular, an incentive structure and accounting methods to encourage users to submit/run their jobs during low-carbon intensity would be useful. Similar to core-hour accounting and budgeting, HPC users should also be provided a carbon budget as a part of their allocation, and they could be prioritized to reduce their queue wait time if the carbon footprint of their jobs have been economical. There is a strong need to design, develop, and deploy carbon-intensity-aware job schedulers to exploit these opportunities across geographically distributed HPC centers. Currently, we already have some CloudBank-related programs which allow users to submit their jobs to different centers. However, robust system software support for real-time and automatic distribution of jobs is needed.
\end{mybox}

%% file: sections/part3.tex
\section{Operational and Embodied Carbon}
\label{sec:p3}

In previous sections, we have addressed two key aspects of sustainability in HPC centers: embodied carbon in supercomputer components and operational carbon, which is influenced by regional carbon intensity. \textit{In this section, we aim to integrate these factors and examine the total carbon footprint of supercomputer upgrades. By taking a holistic view of embodied and operational carbon, we can gain a more comprehensive understanding of the environmental impact of these upgrades.}

\begin{mybox}{blue}{blue}
\textbf{RQ \showRQcounter.} \textit{What are carbon footprint and performance trade-offs as supercomputing facilities consider hardware upgrades, esp. multi-generation GPU upgrades? Will the introduction of embodied carbon due to upgrade be offset by savings in operational carbon footprint due to more energy-efficient newer generation hardware? Do these trade-offs depend on the ``greenness'' of the energy mix?}
\end{mybox}

\begin{table}
\small
\centering
\caption{Different generations of nodes analyzed.}
\vspace{-4mm}
\scalebox{1.0}{
\begin{tabular}{lll}
\toprule
\textbf{Name} & \textbf{GPU} & \textbf{CPU} \\ 
\midrule
\midrule
P100 & 4$\times$ NVIDIA Tesla P100 PCIe & 2$\times$ Intel Xeon CPU E5-2680 \\
\midrule
V100 & 4$\times$ NVIDIA Tesla V100 SXM2 & 2$\times$ Intel Xeon Gold 6240R \\
\midrule
A100 & 4$\times$ NVIDIA A100 PCIe 40GB & 4$\times$ AMD EPYC 7542 \\
\bottomrule
\end{tabular}}
\label{table:part3_nodes}
\vspace{-4mm}
\end{table}

\begin{table}
\small
\centering
\caption{Performance improvement from the node upgrade.}
\vspace{-4mm}
\scalebox{1.0}{
\begin{tabular}{lllll}
\toprule
\makecell[l]{\textbf{Upgrade}\\\textbf{Option}} & \makecell[l]{\textbf{NLP}\\\textbf{Improv.}} &\makecell[l]{\textbf{Vision}\\\textbf{Improv.}} & \makecell[l]{\textbf{CANDLE}\\\textbf{Improv.}} & \makecell[l]{\textbf{Average}\\\textbf{Improv.}} \\ 
\midrule
\midrule
P100 to V100 & 44.4\% & 41.2\% & 45.5\% & 43.4\% \\
\midrule
P100 to A100 & 59.0\% & 60.2\% & 68.3\% & 62.5\% \\
\midrule
V100 to A100 & 25.6\% & 35.8\% & 44.4\% & 35.9\% \\
\bottomrule
\end{tabular}}
\label{table:part3_perf}
\vspace{-4mm}
\end{table}

\noindent\textbf{Result and Analysis.} To investigate the impact of system upgrades on performance and carbon footprint, we perform benchmarking on three nodes of different generations, denoted as P100, V100, and A100 nodes shown in Table~\ref{table:part3_nodes}. These three generations represent NVIDIA's three major datacenter GPU architectures released in the past: Pascal, Volta, and Ampere.  

We clarify that these experiments and analyses are primarily based on GPUs for simplicity, as they are likely to be among the most dominant contributing factors toward the overall carbon footprint of the data center (embodied and operational).

\begin{figure}[t]
    \centering
    \includegraphics[scale=0.354]{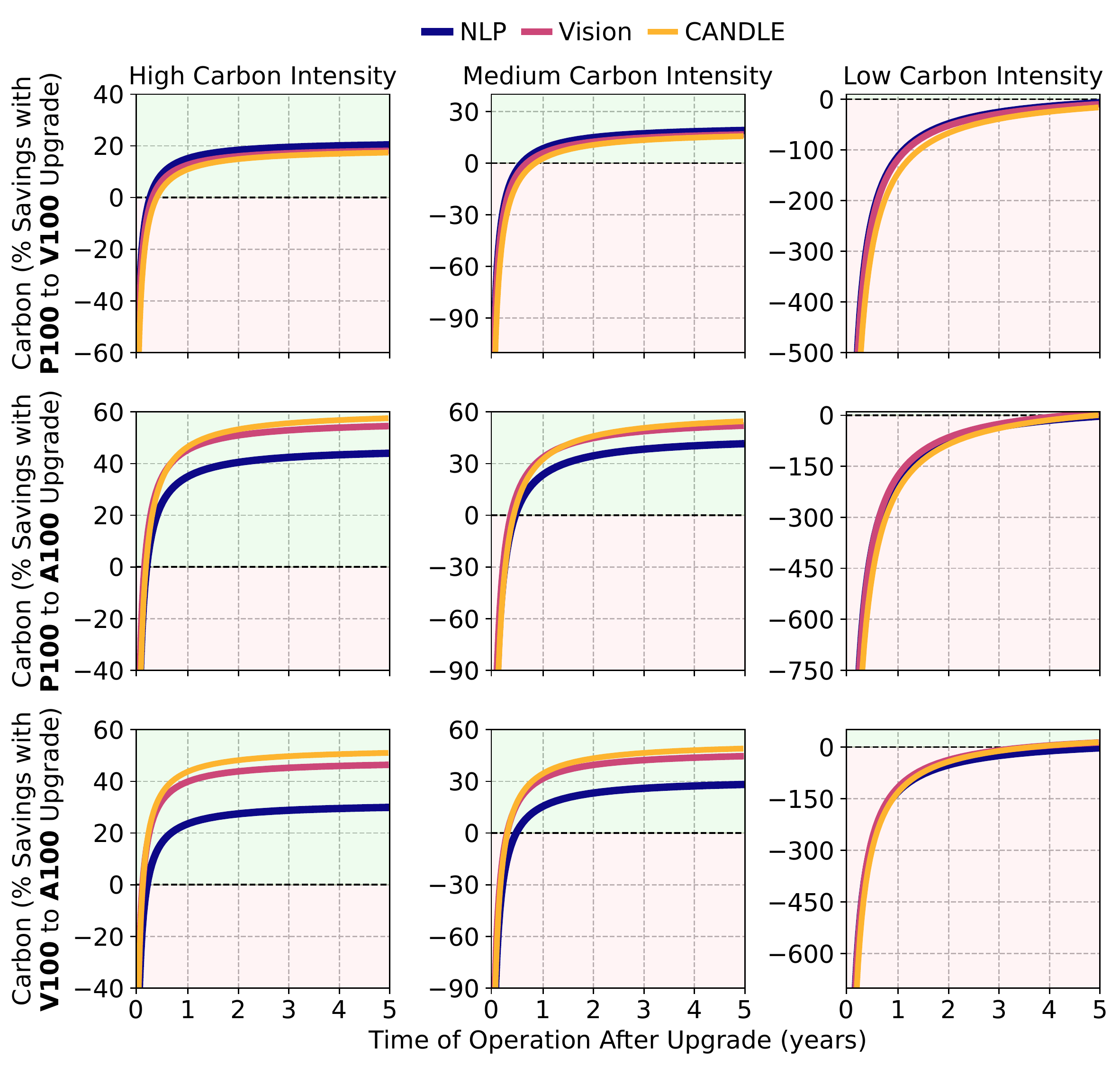}
    \vspace{-0.8cm}
    \caption{By upgrading a GPU system, one can reduce carbon emission over time despite a large carbon footprint increase initially. The amortization rate depends on the region's carbon intensity.}
    \label{fig:carbon_with_time}
    \vspace{-8mm}
\end{figure}

Our evaluation considers three upgrade options between node generations: P100 to V100, P100 to A100, and V100 to A100. To assess performance improvements, we conducted benchmarks listed in Table~\ref{table:benchmarks} and recorded the results for each benchmark set in Table~\ref{table:part3_perf}. The data indicates that all upgrade options delivered notable performance improvements, as expected, with the largest gains when upgrading from P100 to A100 due to a longer gap in time between those generations. The performance improvements ranged from 25\% to almost 70\%. Notably, the CANDLE benchmark demonstrated greater performance improvements than the other two benchmarks across all three upgrade options. Overall, the performance gains from the upgrades were significant. In the following section, we explore the environmental impact of the upgrades in terms of their carbon footprint and make a case that performance improvements alone may not sufficient for considering the upgrade decision.

The figure shown in Fig.~\ref{fig:carbon_with_time} presents the results of our analysis of three upgrade scenarios, each represented by a row (P100 to V100, P100 to A100, and V100 to A100), and different levels of average carbon intensity, represented by columns (high, medium, and low). Three lines on each subplot represent three different workloads (NLP, Vision, and CANDLE). 

We evaluated the carbon footprint savings achieved by upgrading the system over a five-year period following the node upgrade. The red region indicates when the upgraded option resulted in a higher carbon footprint than not upgrading, while the green region indicates the upgrade has resulted in carbon footprint savings. 

First, we analyze the first column (that is, the carbon intensity is held constant) of Fig.~\ref{fig:carbon_with_time}. As expected, all curves start from a negative point because an upgrade immediately incurs embodied carbon cost, and it takes some time before this ``tax'' can be paid. Along the way, it is offset is by saving operational energy over time (newer hardware is typically more energy efficient and hence, results in lower energy consumption). This is why almost all curves go toward the ``green'' region, albeit at a different rate. 


The rate/steepness of the curve depends on the workload (NLP vs. CANDLE) and upgrade tier (P100 to A100 vs V100 to A100). In general, the energy efficiency improvements are the highest when upgrading from P100 to A100, and hence, the embodied carbon ``tax'' is paid quickly. NLP curve is typically below other Vision and CANDLE workloads because NLP receives the least performance improvement, and hence, the least energy improvement. 

These findings align with the conventional wisdom that when newer, faster, and more energy-efficient hardware is available, we should upgrade the system. Our results partially support that from a carbon-consciousness aspect too, albeit with a caveat that the upgrades cannot be too fast and the window before the tax is offset can vary depending upon the workload being run in the system.

Next, we analyze the effect of carbon intensity on the same decisions. Across columns in Fig.~\ref{fig:carbon_with_time}, we evaluate the carbon footprint reduction in three different carbon intensities: high intensity with an average of 400 g\coo{}/kWh, medium intensity with an average of 200 g\coo{}/kWh, and low intensity with an average of 20 g\coo{}/kWh which is the carbon intensity of hydropower~\cite{gupta2022act}. 

\textit{We make an interesting observation that hardware upgrade benefits can heavily depend on the energy source of the HPC center.} At high carbon intensity, it takes less than half a year to amortize the embodied carbon incurred at system upgrade; at medium carbon intensity, it takes less than a year to amortize the embodied carbon; but at low carbon intensity when the energy source is highly renewable, the amortization time is about five years or more. 

Overall, upgrading is beneficial in terms of performance, but the carbon footprint perspective needs more consideration in the carbon intensity and expected system service life. In regions with high carbon intensity, upgrades can happen when the new generation is released since the new system will quickly amortize its embodied carbon. In regions with an abundant amount of green energy, upgrading would be carbon-friendly only if the system is expected to serve for at least five years approximately. We note that these periods are dependent on the embodied carbon and energy consumption of these devices, and hence, they should be interpreted in the context of the modeled situation. 


\begin{mybox}{green}{green}
\textbf{Insight \showIQcounter.} Hardware upgrades are always attractive due to significant performance improvements, but the upgrade introduces significant embodied carbon which may not be offset quickly -- esp. if the center already runs primarily on renewable energy sources, as could be the case in the future for many centers. In such cases, extending the hardware lifetime could be a worthy option. If the energy source is less green, a quicker upgrade may be desirable. Although we note that there are many other indirect side effects associated with the upgrades, including monetary investment, potential redesign of the data center, etc. which may further increase the carbon footprint. 
\end{mybox}

\begin{figure}[t]
    \centering
    \includegraphics[scale=0.354]{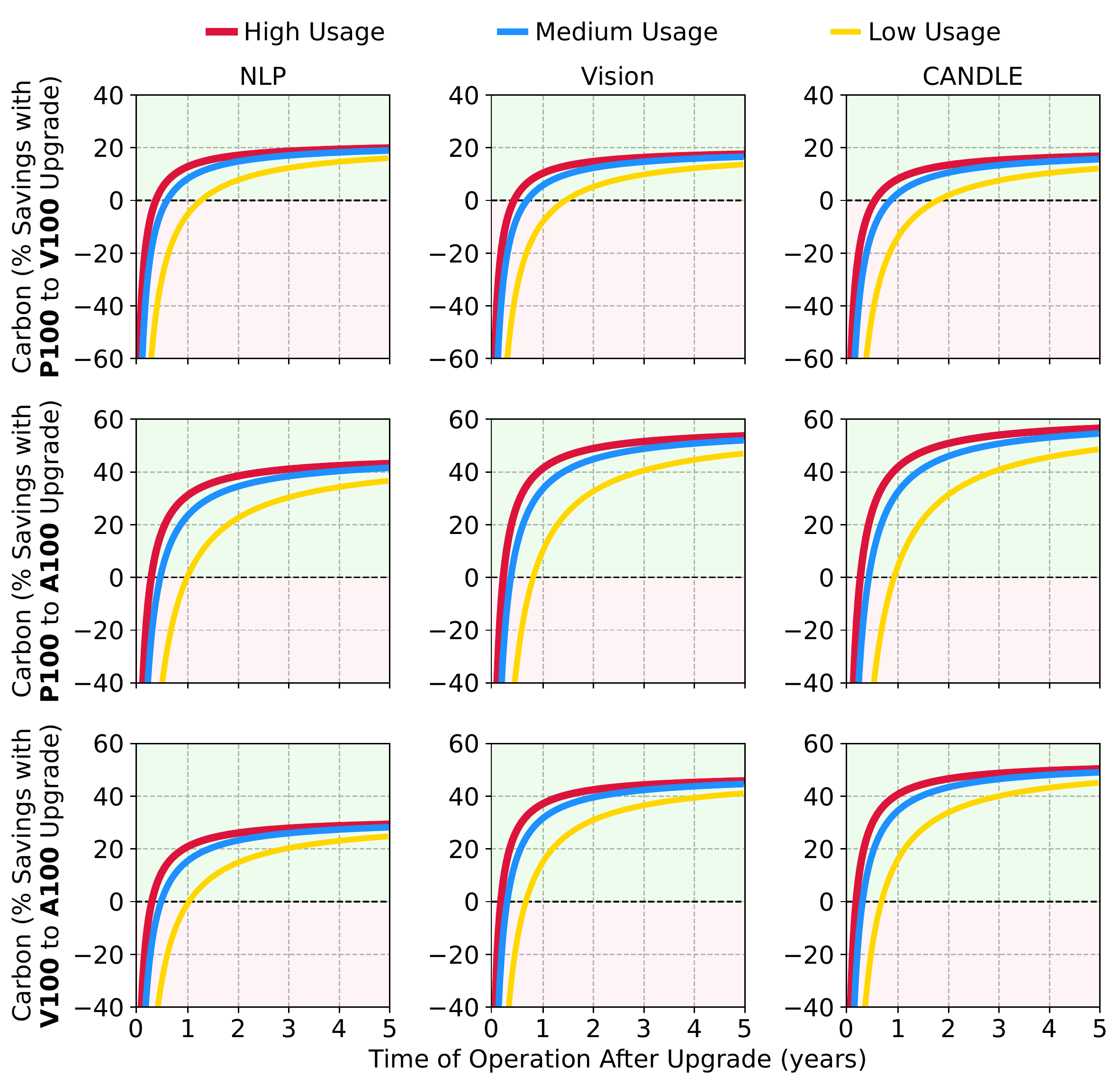}
    \vspace{-0.8cm}
    \caption{Carbon saving curve after the GPU system upgrade under different usage patterns.}
    \vspace{-6mm}
    \label{fig:carbon_diff_util}

\end{figure}

\begin{mybox}{blue}{blue}
\textbf{RQ \showRQcounter.} \textit{Does carbon footprint trade-offs at supercomputing facilities consider hardware upgrades depending upon the average load or utilization of the current system, and the workload being run?}
\end{mybox}

\noindent\textbf{Result and Analysis.} HPC centers have high utilization levels, but not all GPUs are utilized at all times. In fact, multiple HPC and data centers have reported low GPU utilization levels~\cite{weng2022mlaas,li2022ai,jeon2019analysis}. Therefore, we investigate this aspect too. 

In Fig.~\ref{fig:carbon_diff_util}, we keep the average carbon intensity constant at 200 g\coo{}/kWh and vary the average GPU usage rate of the nodes, which represents the percentage of time the GPU is being used. The three carbon-saving curves in each subplot represent three different GPU utilization levels. We first assume that all the GPU nodes are allocated by users 100\% of the time, and choose 40\% GPU usage as the medium usage to align with a production trace~\cite{weng2022mlaas,li2022ai,jeon2019analysis}. The high and low usage numbers are selected based on 1.5$\times$ more and less of the medium usage, respectively. Our previous analysis in Fig.~\ref{fig:carbon_with_time} is conducted with medium usage.

In Fig.~\ref{fig:carbon_diff_util}, we show that depending on the user usage pattern, the time it takes to amortize the embodied carbon varies. The difference is not as significant as the carbon intensity, where it can be multiple years of difference. Nevertheless, it would still substantially impact the saved carbon footprint at a certain point in time after the upgrade. Using the NLP benchmark set as an example, when we upgrade from the V100 system to the A100 system, after one year, a high/medium usage pattern would result in approximately 20\% carbon footprint reduction, whereas the low usage pattern has just paid off the initial embodied carbon of the upgrade.

\begin{mybox}{green}{green}
\textbf{Insight \showIQcounter.} If the center has limited GPU utilization, extending the hardware lifetime could be a worthy option, but it also depends upon the mix of workloads being run and their energy consumption characteristics. If the GPU utilization is high, a quicker upgrade may be desirable, since the savings in operational carbon footprint could quickly offset the introduction of embodied carbon. Nevertheless, this decision is still heavily influenced by the ``greenness'' of the energy source.
\end{mybox}

\begin{mybox}{yellow}{yellow}
\textbf{Implication.} The HPC centers should consider carbon footprint in their hardware upgrade decisions. We recognize that carbon awareness alone is not the determining factor -- traditionally, cost and performance improvements have dictated the timing of such decisions. However, as carbon net-zero aims become more commonplace, the centers should have methods, such as those introduced in this paper, to evaluate the lifetime of a hardware generation and if extending it would be useful. The HPC centers need to continuously monitor and evaluate the carbon-intensity and GPU utilization load to determine when to upgrade because carbon-intensity and GPU utilization are among the most dominant factor in the hardware-upgrade decision-making which attempts to minimize carbon footprint. Assessing the carbon-friendliness of HPC centers involves considering several factors. These factors include the initial embodied carbon footprint, the operational carbon that is linked to the regional carbon intensity and user usage pattern, and the expected operating lifetime of the HPC system. These factors are not standard yet. Large-scale HPC applications would have a large operational carbon footprint due to the heavy computation carried out across multiple nodes. 
\end{mybox}



\section{Threats To Validity and Discussion}
\label{sec:threats}


We recognize and acknowledge that our analysis is not immune to threats to validity. For instance, while our current analysis models the embodied carbon footprint of the hardware and operational carbon footprint, it does not account for carbon emissions related to other side effects of upgrading the system. Such emissions arise not only during manufacturing and packaging but also during transportation, installation, and recycling. System upgrades often require increasing building capacity, changes in cooling infrastructure, additional staffing, and acquisition of other compute/storage components to match the increase in the compute capacity -- these aspects lead to additional carbon emission. Hardware cost is also an important aspect, but we do not explicitly include it because hardware cost may depend on many factors and vary across geographical regions, the scale of the system, and may change over time. Unfortunately, as a community, we are at a very early stage to have sufficiently detailed and accurate modeling of these by-product effects. We are hoping that this work will spur a new research direction in the HPC community to standardize and document these aspects -- which have not been central for performance benchmarking, but will be critical for HPC carbon benchmarking. 

We acknowledge that fab yield differs across manufacturers and lithography. Due to the lack of a standardized public database providing accurate yield data, we configured it to be based on prior works. Similarly, the PUE metric, while challenging to estimate with seasonal variation, can be approximated well with IT and cooling energy monitors available in most HPC datacenters. The operational carbon can be calculated using Eq.~\ref{eq:operational} (e.g., energy from power measurement tools (e.g., NVML, RAPL) multiplied by the carbon intensity). Note that the embodied carbon relative to the operational carbon varies across P100, V100, and A100 GPUs. 

Our choice to select the three workloads from a broad range of HPC-relevant benchmarks was because they cover a diverse range of resource consumption characteristics, resulting in different trends that capture the spectrum. Adding more benchmarks makes the interpretation and presentation complex, but they should be considered in the future as they can lead to more insights, esp. given that HPC workloads are rapidly evolving with new AI and quantum use cases. The operational carbon footprint of large-scale applications is also an important consideration and potential threat to validity, but these can be performed in-house by different HPC systems as they may not always make their representative workloads publicly available. The key actions include but are not limited to, measuring the carbon footprint of large-scale applications on different computing hardware and making carbon-aware HPC scheduling and hardware upgrade decisions based on geographical carbon intensity and application characteristics.  The HPC site may have a contract with a local power provider which needs to be considered when accounting for the carbon footprint. Additionally, migrating users to new architectures (e.g., GPUs) also incurs an indirect carbon footprint and has trade-offs in terms of performance, esp. if the system is not being procured for a high FLOP count but instead for running a specific workload. These aspects are not explicitly accounted in this study.
 
The popular energy efficiency metric (FLOPS/Watt) alone does not fully capture the environmental impact of an HPC system because it does not capture embodied carbon and carbon intensity of the energy source. A system with higher energy efficiency does not necessarily mean it has lower operational carbon footprint – because operational carbon also depends on the carbon intensity of the energy being fed (e.g., operation of system A (20GFLOPS/Watts) may be ``greener'' than B (50GFLOPS/Watts) if A uses hydropower while B uses gas).

We recognize that cost is an important consideration during hardware upgrades. For example, some low-cost systems can indeed have a carbon lower footprint as well, depending upon the load and carbon intensity. However, hardware cost may depend on many factors and vary (across geographical regions, scale of the system, and may change over time, etc.), hence integrating it into the study requires careful consideration with restrictive assumptions.

%% file: sections/related_work.tex
\section{Related Work}
\label{sec:relat}

\noindent\textbf{Study of HPC system components. } Various works have addressed the energy efficiency of CPU~\cite{papadimitriou2017harnessing,yuan2006energy,podzimek2015analyzing}, DRAM~\cite{o2017fine,giridhar2013exploring,hassan2019crow}, and storage components~\cite{tomes2017comparative,park2011comprehensive,harris2020ultra} in a HPC system.
GPUs have been widely integrated into today's HPC systems to accelerate deep learning applications, hence a variety of efforts have focused on improving the GPU device energy efficiency~\cite{arunkumar2019understanding,majumdar2017dynamic,kandiah2021accelwattch,leng2013gpuwattch,sethia2014equalizer,khairy2020locality}. These works have only focused on the energy efficiency of individual components, not on the carbon emission during manufacturing and operation.

\vspace{1.mm}
\noindent\textbf{Green computing in large-scale GPU systems. } The environmental impact of HPC systems has been an important research topic in the past. GPU-NEST~\cite{jahanshahi2020gpu} has examined the power and performance behaviors of multi-GPU inference systems. 
Helios~\cite{hu2021characterization} proposed a cluster energy-saving service for GPU datacenter running deep learning workloads.  Patki et. al.~\cite{patki2019comparing} applied GPU power and frequency capping in a scientific workflow to improve power efficiency while preserving performance. Our work goes beyond the investigation of energy consumption: we analyze the regional carbon intensity to relate the energy consumption to the actual carbon emission. Various works have investigated the operational carbon footprint when running workloads such as bioinformatics and astrophysics in large-scale systems~\cite{grealey2022carbon,portegies2020ecological,wu2022sustainable,patterson2021carbon}. Our work is distinct from the previous ones as we combine the embodied carbon footprint, operational carbon footprint, regional carbon intensity, performance, system service life, and usage pattern to conduct a holistic analysis. This study is complementary to various emerging energy-efficient HPC network topologies~\cite{andujar2023energy,georgakoudis2019evaluating}.

\vspace{1.mm}
\noindent\textbf{Carbon footprint modeling. } Wang~\cite{wang2008meeting} and Totally Green~\cite{chang2012totally} proposed to take the production-operation-recycling product life cycle into sustainability analysis. In regard to the production perspective, ACT~\cite{gupta2022act} released an embodied carbon footprint modeling tool that is useful for conducting embodied carbon analysis for mobile devices. Our work is not limited to embodied carbon modeling and we also model the operational carbon on systems of different generations under different regional carbon intensities and usage patterns for HPC systems. Chien et. al.~\cite{dietrich2022navigating} developed a model for historical NSF XSEDE systems to examine their future trends, which complements our work with extra analysis of capital, total cost of ownership, power, and more. Overall, this paper is the first work to provide carbon modeling methods and tools for HPC practitioners and identify various areas of opportunities for HPC system facilities. 

\section{Conclusion}
\label{sec:conclud}

We present a carbon footprint analysis framework that addresses a series of research questions related to HPC system sustainability. We have conducted modeling of the embodied carbon footprint of HPC system components, investigations into how regional carbon intensity affects the system, and end-to-end characterizations of the carbon footprint throughout the system life cycle. We hope that our methodology, framework, and study encourage HPC researchers to promote sustainable HPC and promote more research efforts toward carbon neutrality in the community.

%% file: main.bbl
\begin{thebibliography}{90}
\providecommand{\natexlab}[1]{#1}
\providecommand{\url}[1]{\texttt{#1}}
\expandafter\ifx\csname urlstyle\endcsname\relax
  \providecommand{\doi}[1]{doi: #1}\else
  \providecommand{\doi}{doi: \begingroup \urlstyle{rm}\Url}\fi

\bibitem[Portegies~Zwart(2020)]{portegies2020ecological}
Simon Portegies~Zwart.
\newblock The ecological impact of high-performance computing in astrophysics.
\newblock \emph{Nature Astronomy}, 4\penalty0 (9):\penalty0 819--822, 2020.

\bibitem[Panda et~al.(2021)Panda, Subramoni, Chu, and
  Bayatpour]{panda2021mvapich}
Dhabaleswar~Kumar Panda, Hari Subramoni, Ching-Hsiang Chu, and Mohammadreza
  Bayatpour.
\newblock The mvapich project: Transforming research into high-performance mpi
  library for hpc community.
\newblock \emph{Journal of Computational Science}, 52:\penalty0 101208, 2021.

\bibitem[Feichtinger et~al.(2011)Feichtinger, Donath, K{\"o}stler, G{\"o}tz,
  and R{\"u}de]{feichtinger2011walberla}
Christian Feichtinger, Stefan Donath, Harald K{\"o}stler, Jan G{\"o}tz, and
  Ulrich R{\"u}de.
\newblock Walberla: Hpc software design for computational engineering
  simulations.
\newblock \emph{Journal of Computational Science}, 2\penalty0 (2):\penalty0
  105--112, 2011.

\bibitem[Commission(2021)]{usitc}
United States International~Trade Commission.
\newblock Data centers around the world: A quick look, 2021.
\newblock URL
  \url{https://www.usitc.gov/publications/332/executive_briefings/ebot_data_centers_around_the_world.pdf}.

\bibitem[Statista(2023)]{statista}
Statista.
\newblock Data centers - statistics \& facts, 2023.
\newblock URL \url{https://www.statista.com/topics/6165/data-centers/}.

\bibitem[Laboratory(2023)]{frontier_arch}
Oak Ridge~National Laboratory.
\newblock Frontier's architecture, 2023.
\newblock URL
  \url{https://olcf.ornl.gov/wp-content/uploads/Frontiers-Architecture-Frontier-Training-Series-final.pdf}.

\bibitem[Gupta et~al.(2022{\natexlab{a}})Gupta, Elgamal, Hills, Wei, Lee,
  Brooks, and Wu]{gupta2022act}
Udit Gupta, Mariam Elgamal, Gage Hills, Gu-Yeon Wei, Hsien-Hsin~S Lee, David
  Brooks, and Carole-Jean Wu.
\newblock Act: Designing sustainable computer systems with an architectural
  carbon modeling tool.
\newblock In \emph{Proceedings of the 49th Annual International Symposium on
  Computer Architecture}, pages 784--799, 2022{\natexlab{a}}.

\bibitem[Chang et~al.(2012)Chang, Meza, Ranganathan, Shah, Shih, and
  Bash]{chang2012totally}
Jichuan Chang, Justin Meza, Parthasarathy Ranganathan, Amip Shah, Rocky Shih,
  and Cullen Bash.
\newblock Totally green: evaluating and designing servers for lifecycle
  environmental impact.
\newblock \emph{ACM SIGPLAN Notices}, 47\penalty0 (4):\penalty0 25--36, 2012.

\bibitem[Gupta et~al.(2022{\natexlab{b}})Gupta, Kim, Lee, Tse, Lee, Wei,
  Brooks, and Wu]{gupta2022chasing}
Udit Gupta, Young~Geun Kim, Sylvia Lee, Jordan Tse, Hsien-Hsin~S Lee, Gu-Yeon
  Wei, David Brooks, and Carole-Jean Wu.
\newblock Chasing carbon: The elusive environmental footprint of computing.
\newblock \emph{IEEE Micro}, 42\penalty0 (4):\penalty0 37--47,
  2022{\natexlab{b}}.

\bibitem[Andrae and Edler(2015)]{andrae2015global}
Anders~SG Andrae and Tomas Edler.
\newblock On global electricity usage of communication technology: trends to
  2030.
\newblock \emph{Challenges}, 6\penalty0 (1):\penalty0 117--157, 2015.

\bibitem[Amazon(2021)]{amazon_netzero}
Amazon.
\newblock Amazon sustainability report, 2021.
\newblock URL
  \url{https://sustainability.aboutamazon.com/2021-sustainability-report.pdf}.

\bibitem[Google(2022)]{google_netzero}
Google.
\newblock Google environmental report, 2022.
\newblock URL
  \url{https://www.gstatic.com/gumdrop/sustainability/google-2022-environmental-report.pdf}.

\bibitem[Meta(2021)]{meta_netzero}
Meta.
\newblock Meta sustainability report, 2021.
\newblock URL
  \url{https://sustainability.fb.com/wp-content/uploads/2022/06/Meta-2021-Sustainability-Report.pdf}.

\bibitem[Apple(2022)]{apple_netzero}
Apple.
\newblock Environmental progress report, 2022.
\newblock URL
  \url{https://www.apple.com/environment/pdf/Apple_Environmental_Progress_Report_2022.pdf}.

\bibitem[Cao et~al.(2022)Cao, Zhou, Hu, Wang, and Wen]{cao2022towards}
Zhiwei Cao, Xin Zhou, Han Hu, Zhi Wang, and Yonggang Wen.
\newblock Towards a systematic survey for carbon neutral data centers.
\newblock \emph{IEEE Communications Surveys \& Tutorials}, 2022.

\bibitem[Acun et~al.(2023)Acun, Lee, Kazhamiaka, Maeng, Gupta, Chakkaravarthy,
  Brooks, and Wu]{acun2023carbon}
Bilge Acun, Benjamin Lee, Fiodar Kazhamiaka, Kiwan Maeng, Udit Gupta, Manoj
  Chakkaravarthy, David Brooks, and Carole-Jean Wu.
\newblock Carbon explorer: A holistic framework for designing carbon aware
  datacenters.
\newblock In \emph{Proceedings of the 28th ACM International Conference on
  Architectural Support for Programming Languages and Operating Systems, Volume
  2}, pages 118--132, 2023.

\bibitem[Gu and Budati(2020)]{gu2020energy}
Yi~Gu and Chandu Budati.
\newblock Energy-aware workflow scheduling and optimization in clouds using bat
  algorithm.
\newblock \emph{Future Generation Computer Systems}, 113:\penalty0 106--112,
  2020.

\bibitem[Li et~al.(2023)Li, Samsi, Gadepally, and Tiwari]{li2023green}
Baolin Li, Siddharth Samsi, Vijay Gadepally, and Devesh Tiwari.
\newblock Green carbon footprint for model inference serving via exploiting
  mixed-quality models and gpu partitioning.
\newblock \emph{arXiv preprint arXiv:2304.09781}, 2023.

\bibitem[Wang(2008)]{wang2008meeting}
David Wang.
\newblock Meeting green computing challenges.
\newblock In \emph{2008 10th Electronics Packaging Technology Conference},
  pages 121--126. IEEE, 2008.

\bibitem[Jones et~al.(2013)Jones, Chen, Collinge, Xu, Schaefer, Landis, and
  Bilec]{jones2013considering}
Alex~K Jones, Yiran Chen, William~O Collinge, Haifeng Xu, Laura~A Schaefer,
  Amy~E Landis, and Melissa~M Bilec.
\newblock Considering fabrication in sustainable computing.
\newblock In \emph{2013 IEEE/ACM International Conference on Computer-Aided
  Design (ICCAD)}, pages 206--210. IEEE, 2013.

\bibitem[Kline et~al.(2016)Kline, Parshook, Ge, Brunvand, Melhem, Chrysanthis,
  and Jones]{kline2016holistically}
Donald Kline, Nikolas Parshook, Xiaoyu Ge, Erik Brunvand, Rami Melhem, Panos~K
  Chrysanthis, and Alex~K Jones.
\newblock Holistically evaluating the environmental impacts in modern computing
  systems.
\newblock In \emph{2016 Seventh International Green and Sustainable Computing
  Conference (IGSC)}, pages 1--8. IEEE, 2016.

\bibitem[Kline~Jr et~al.(2019)Kline~Jr, Parshook, Ge, Brunvand, Melhem,
  Chrysanthis, and Jones]{kline2019greenchip}
Donald Kline~Jr, Nikolas Parshook, Xiaoyu Ge, Erik Brunvand, Rami Melhem,
  Panos~K Chrysanthis, and Alex~K Jones.
\newblock Greenchip: A tool for evaluating holistic sustainability of modern
  computing systems.
\newblock \emph{Sustainable Computing: Informatics and Systems}, 22:\penalty0
  322--332, 2019.

\bibitem[SPIL(2020)]{spil}
SPIL.
\newblock Corporate social responsibility report, 2020.
\newblock URL \url{https://www.spil.com.tw/Files/pdf-en/2019-en.pdf}.

\bibitem[Seagate(2023)]{seagate}
Seagate.
\newblock Seagate product sustainability, 2023.
\newblock URL
  \url{https://www.seagate.com/global-citizenship/product-sustainability/}.

\bibitem[List(2022{\natexlab{a}})]{top500}
Top-500 List.
\newblock {November 2022}, 2022{\natexlab{a}}.
\newblock URL \url{https://www.top500.org/lists/top500/2022/11/}.

\bibitem[Report(2022)]{hynix}
SK~Hynix~Sustainability Report, 2022.
\newblock URL \url{https://www.skhynix.com/sustainability/UI-FR-SA1601/}.

\bibitem[Labratory(2023)]{frontier}
Oak Ridge~National Labratory.
\newblock {Frontierr}, 2023.
\newblock URL \url{https://www.olcf.ornl.gov/frontier/}.

\bibitem[Consortium(2023)]{LUMI}
LUMI Consortium.
\newblock {LUMI Supercomputer}, 2023.
\newblock URL \url{https://lumi-supercomputer.eu/}.

\bibitem[NERSC(2023)]{perlmutter}
NERSC.
\newblock {Perlmutter: High Performance Computing Optimized for Science}, 2023.
\newblock URL \url{https://perlmutter.carrd.co/}.

\bibitem[Anthony et~al.(2020)Anthony, Kanding, and
  Selvan]{anthony2020carbontracker}
Lasse F~Wolff Anthony, Benjamin Kanding, and Raghavendra Selvan.
\newblock Carbontracker: Tracking and predicting the carbon footprint of
  training deep learning models.
\newblock \emph{arXiv preprint arXiv:2007.03051}, 2020.

\bibitem[Company(2023{\natexlab{a}})]{kn}
Kansai Electric~Power Company.
\newblock {Kansai Electric Power}, 2023{\natexlab{a}}.
\newblock URL \url{https://www.kepco.co.jp/english/}.

\bibitem[Company(2023{\natexlab{b}})]{tk}
Tokyo Electric~Power Company.
\newblock {TEPCO Power Grid}, 2023{\natexlab{b}}.
\newblock URL \url{https://www.tepco.co.jp/en/pg/index-e.html}.

\bibitem[ESO(2023{\natexlab{a}})]{eso}
National~Grid ESO.
\newblock {Welcome to ESO}, 2023{\natexlab{a}}.
\newblock URL \url{https://www.nationalgrideso.com/}.

\bibitem[ISO(2023)]{ciso}
California ISO.
\newblock {California Independent System Operator}, 2023.
\newblock URL \url{https://www.caiso.com/Pages/default.aspx}.

\bibitem[PJM(2023)]{pjm}
PJM.
\newblock {PJM Interconnection}, 2023.
\newblock URL \url{https://www.pjm.com/}.

\bibitem[MISO(2023)]{miso}
MISO.
\newblock {Midcontinent Independent System Operator}, 2023.
\newblock URL \url{https://www.misoenergy.org/about/}.

\bibitem[ERCOT(2023)]{ercot}
ERCOT.
\newblock {Electric Reliability Council of Texas}, 2023.
\newblock URL \url{https://www.ercot.com/}.

\bibitem[ESO(2023{\natexlab{b}})]{eso_ci}
National~Grid ESO.
\newblock Carbon intensity api, 2023{\natexlab{b}}.
\newblock URL \url{https://carbonintensity.org.uk/}.

\bibitem[Maps(2023)]{electricity_map}
Electricity Maps.
\newblock Reduce carbon emissions with actionable electricity data, 2023.
\newblock URL \url{https://www.electricitymaps.com/}.

\bibitem[Devlin et~al.(2018)Devlin, Chang, Lee, and Toutanova]{devlin2018bert}
Jacob Devlin, Ming-Wei Chang, Kenton Lee, and Kristina Toutanova.
\newblock Bert: Pre-training of deep bidirectional transformers for language
  understanding.
\newblock \emph{arXiv preprint arXiv:1810.04805}, 2018.

\bibitem[Sanh et~al.(2019)Sanh, Debut, Chaumond, and Wolf]{sanh2019distilbert}
Victor Sanh, Lysandre Debut, Julien Chaumond, and Thomas Wolf.
\newblock Distilbert, a distilled version of bert: smaller, faster, cheaper and
  lighter.
\newblock \emph{arXiv preprint arXiv:1910.01108}, 2019.

\bibitem[Song et~al.(2020)Song, Tan, Qin, Lu, and Liu]{song2020mpnet}
Kaitao Song, Xu~Tan, Tao Qin, Jianfeng Lu, and Tie-Yan Liu.
\newblock Mpnet: Masked and permuted pre-training for language understanding.
\newblock \emph{Advances in Neural Information Processing Systems},
  33:\penalty0 16857--16867, 2020.

\bibitem[Liu et~al.(2019)Liu, Ott, Goyal, Du, Joshi, Chen, Levy, Lewis,
  Zettlemoyer, and Stoyanov]{liu2019roberta}
Yinhan Liu, Myle Ott, Naman Goyal, Jingfei Du, Mandar Joshi, Danqi Chen, Omer
  Levy, Mike Lewis, Luke Zettlemoyer, and Veselin Stoyanov.
\newblock Roberta: A robustly optimized bert pretraining approach.
\newblock \emph{arXiv preprint arXiv:1907.11692}, 2019.

\bibitem[Lewis et~al.(2019)Lewis, Liu, Goyal, Ghazvininejad, Mohamed, Levy,
  Stoyanov, and Zettlemoyer]{lewis2019bart}
Mike Lewis, Yinhan Liu, Naman Goyal, Marjan Ghazvininejad, Abdelrahman Mohamed,
  Omer Levy, Ves Stoyanov, and Luke Zettlemoyer.
\newblock Bart: Denoising sequence-to-sequence pre-training for natural
  language generation, translation, and comprehension.
\newblock \emph{arXiv preprint arXiv:1910.13461}, 2019.

\bibitem[He et~al.(2016)He, Zhang, Ren, and Sun]{he2016deep}
Kaiming He, Xiangyu Zhang, Shaoqing Ren, and Jian Sun.
\newblock Deep residual learning for image recognition.
\newblock In \emph{Proceedings of the IEEE conference on computer vision and
  pattern recognition}, pages 770--778, 2016.

\bibitem[Xie et~al.(2017)Xie, Girshick, Doll{\'a}r, Tu, and
  He]{xie2017aggregated}
Saining Xie, Ross Girshick, Piotr Doll{\'a}r, Zhuowen Tu, and Kaiming He.
\newblock Aggregated residual transformations for deep neural networks.
\newblock In \emph{Proceedings of the IEEE conference on computer vision and
  pattern recognition}, pages 1492--1500, 2017.

\bibitem[Ma et~al.(2018)Ma, Zhang, Zheng, and Sun]{ma2018shufflenet}
Ningning Ma, Xiangyu Zhang, Hai-Tao Zheng, and Jian Sun.
\newblock Shufflenet v2: Practical guidelines for efficient cnn architecture
  design.
\newblock In \emph{Proceedings of the European conference on computer vision
  (ECCV)}, pages 116--131, 2018.

\bibitem[Simonyan and Zisserman(2014)]{simonyan2014very}
Karen Simonyan and Andrew Zisserman.
\newblock Very deep convolutional networks for large-scale image recognition.
\newblock \emph{arXiv preprint arXiv:1409.1556}, 2014.

\bibitem[Mao et~al.(2022)Mao, Qi, Chen, Li, Duan, Ye, He, and
  Xue]{mao2022towards}
Xiaofeng Mao, Gege Qi, Yuefeng Chen, Xiaodan Li, Ranjie Duan, Shaokai Ye, Yuan
  He, and Hui Xue.
\newblock Towards robust vision transformer.
\newblock In \emph{Proceedings of the IEEE/CVF Conference on Computer Vision
  and Pattern Recognition}, pages 12042--12051, 2022.

\bibitem[Can(2022)]{Candle2022}
Ecp-candle.
\newblock \url{https://github.com/ECP-CANDLE/Benchmarks}, 2022.

\bibitem[Wu et~al.(2019)Wu, Taylor, Wozniak, Stevens, Brettin, and
  Xia]{wu2019performance}
Xingfu Wu, Valerie Taylor, Justin~M Wozniak, Rick Stevens, Thomas Brettin, and
  Fangfang Xia.
\newblock Performance, energy, and scalability analysis and improvement of
  parallel cancer deep learning candle benchmarks.
\newblock In \emph{Proceedings of the 48th International Conference on Parallel
  Processing}, pages 1--11, 2019.

\bibitem[AMD(2023)]{amd-mi250x}
AMD.
\newblock {AMD INSTINCT MI200 SERIES ACCELERATOR}, 2023.
\newblock URL
  \url{https://www.amd.com/system/files/documents/amd-instinct-mi200-datasheet.pdf}.

\bibitem[HPE(2023)]{slingshot}
HPE.
\newblock {HPC Slingshot Interconnect}, 2023.
\newblock URL
  \url{https://www.hpe.com/us/en/compute/hpc/slingshot-interconnect.html}.

\bibitem[Braam(2019)]{braam2019lustre}
Peter Braam.
\newblock The lustre storage architecture.
\newblock \emph{arXiv preprint arXiv:1903.01955}, 2019.

\bibitem[Tannu and Nair(2022)]{tannu2022dirty}
Swamit Tannu and Prashant~J Nair.
\newblock The dirty secret of ssds: Embodied carbon.
\newblock \emph{arXiv preprint arXiv:2207.10793}, 2022.

\bibitem[Zuck et~al.(2023)Zuck, Porter, and Tsafrir]{zuck2023degrading}
Aviad Zuck, Donald Porter, and Dan Tsafrir.
\newblock Degrading data to save the planet.
\newblock In \emph{Proceedings of the 19th Workshop on Hot Topics in Operating
  Systems}, pages 61--69, 2023.

\bibitem[Mersy and Krishnan(2023)]{mersy2023toward}
Gabriel Mersy and Sanjay Krishnan.
\newblock Toward a life cycle assessment for the carbon footprint of data.
\newblock In \emph{Proceedings of the 2nd Workshop on Sustainable Computer
  Systems}, pages 1--9, 2023.

\bibitem[Rajbhandari et~al.(2020)Rajbhandari, Rasley, Ruwase, and
  He]{rajbhandari2020zero}
Samyam Rajbhandari, Jeff Rasley, Olatunji Ruwase, and Yuxiong He.
\newblock Zero: Memory optimizations toward training trillion parameter models.
\newblock In \emph{SC20: International Conference for High Performance
  Computing, Networking, Storage and Analysis}, pages 1--16. IEEE, 2020.

\bibitem[Choukse et~al.(2020)Choukse, Sullivan, O’Connor, Erez, Pool,
  Nellans, and Keckler]{choukse2020buddy}
Esha Choukse, Michael~B Sullivan, Mike O’Connor, Mattan Erez, Jeff Pool,
  David Nellans, and Stephen~W Keckler.
\newblock Buddy compression: Enabling larger memory for deep learning and hpc
  workloads on gpus.
\newblock In \emph{2020 ACM/IEEE 47th Annual International Symposium on
  Computer Architecture (ISCA)}, pages 926--939. IEEE, 2020.

\bibitem[Peng et~al.(2020)Peng, Wu, Ren, Li, and Gokhale]{peng2020demystifying}
Ivy Peng, Kai Wu, Jie Ren, Dong Li, and Maya Gokhale.
\newblock Demystifying the performance of hpc scientific applications on
  nvm-based memory systems.
\newblock In \emph{2020 IEEE International Parallel and Distributed Processing
  Symposium (IPDPS)}, pages 916--925. IEEE, 2020.

\bibitem[Ren et~al.(2020)Ren, Wu, and Li]{ren2020exploring}
Jie Ren, Kai Wu, and Dong Li.
\newblock Exploring non-volatility of non-volatile memory for high performance
  computing under failures.
\newblock In \emph{2020 IEEE International Conference on Cluster Computing
  (CLUSTER)}, pages 237--247. IEEE, 2020.

\bibitem[Zhang et~al.(2019)Zhang, Huang, Huang, Xu, and
  Katz]{zhang2019quantifying}
Zhao Zhang, Lei Huang, Ruizhu Huang, Weijia Xu, and Daniel~S Katz.
\newblock Quantifying the impact of memory errors in deep learning.
\newblock In \emph{2019 IEEE International Conference on Cluster Computing
  (CLUSTER)}, pages 1--12. IEEE, 2019.

\bibitem[List(2022{\natexlab{b}})]{green500}
Green-500 List.
\newblock {November 2022}, 2022{\natexlab{b}}.
\newblock URL \url{https://www.top500.org/lists/green500/2022/11/}.

\bibitem[Weng et~al.(2022)Weng, Xiao, Yu, Wang, Wang, He, Li, Zhang, Lin, and
  Ding]{weng2022mlaas}
Qizhen Weng, Wencong Xiao, Yinghao Yu, Wei Wang, Cheng Wang, Jian He, Yong Li,
  Liping Zhang, Wei Lin, and Yu~Ding.
\newblock \text{MLaaS} in the wild: Workload analysis and scheduling in
  large-scale heterogeneous \text{GPU} clusters.
\newblock In \emph{19th USENIX Symposium on Networked Systems Design and
  Implementation (NSDI 22)}, pages 945--960. USENIX Association, 2022.

\bibitem[Li et~al.(2022)Li, Arora, Samsi, Patel, Arcand, Bestor, Byun, Roy,
  Bergeron, Holodnak, et~al.]{li2022ai}
Baolin Li, Rohin Arora, Siddharth Samsi, Tirthak Patel, William Arcand, David
  Bestor, Chansup Byun, Rohan~Basu Roy, Bill Bergeron, John Holodnak, et~al.
\newblock Ai-enabling workloads on large-scale gpu-accelerated system:
  Characterization, opportunities, and implications.
\newblock In \emph{2022 IEEE International Symposium on High-Performance
  Computer Architecture (HPCA)}, pages 1224--1237. IEEE, 2022.

\bibitem[Jeon et~al.(2019)Jeon, Venkataraman, Phanishayee, Qian, Xiao, and
  Yang]{jeon2019analysis}
Myeongjae Jeon, Shivaram Venkataraman, Amar Phanishayee, Junjie Qian, Wencong
  Xiao, and Fan Yang.
\newblock Analysis of large-scale multi-tenant gpu clusters for dnn training
  workloads.
\newblock In \emph{USENIX Annual Technical Conference}, pages 947--960, 2019.

\bibitem[Papadimitriou et~al.(2017)Papadimitriou, Kaliorakis, Chatzidimitriou,
  Gizopoulos, Lawthers, and Das]{papadimitriou2017harnessing}
George Papadimitriou, Manolis Kaliorakis, Athanasios Chatzidimitriou, Dimitris
  Gizopoulos, Peter Lawthers, and Shidhartha Das.
\newblock Harnessing voltage margins for energy efficiency in multicore cpus.
\newblock In \emph{Proceedings of the 50th Annual IEEE/ACM International
  Symposium on Microarchitecture}, pages 503--516, 2017.

\bibitem[Yuan and Nahrstedt(2006)]{yuan2006energy}
Wanghong Yuan and Klara Nahrstedt.
\newblock Energy-efficient cpu scheduling for multimedia applications.
\newblock \emph{ACM Transactions on Computer Systems (TOCS)}, 24\penalty0
  (3):\penalty0 292--331, 2006.

\bibitem[Podzimek et~al.(2015)Podzimek, Bulej, Chen, Binder, and
  Tuma]{podzimek2015analyzing}
Andrej Podzimek, Lubom{\'\i}r Bulej, Lydia~Y Chen, Walter Binder, and Petr
  Tuma.
\newblock Analyzing the impact of cpu pinning and partial cpu loads on
  performance and energy efficiency.
\newblock In \emph{2015 15th IEEE/ACM International Symposium on Cluster, Cloud
  and Grid Computing}, pages 1--10. IEEE, 2015.

\bibitem[O'Connor et~al.(2017)O'Connor, Chatterjee, Lee, Wilson, Agrawal,
  Keckler, and Dally]{o2017fine}
Mike O'Connor, Niladrish Chatterjee, Donghyuk Lee, John Wilson, Aditya Agrawal,
  Stephen~W Keckler, and William~J Dally.
\newblock Fine-grained dram: Energy-efficient dram for extreme bandwidth
  systems.
\newblock In \emph{Proceedings of the 50th Annual IEEE/ACM International
  Symposium on Microarchitecture}, pages 41--54, 2017.

\bibitem[Giridhar et~al.(2013)Giridhar, Cieslak, Duggal, Dreslinski, Chen,
  Patti, Hold, Chakrabarti, Mudge, and Blaauw]{giridhar2013exploring}
Bharan Giridhar, Michael Cieslak, Deepankar Duggal, Ronald Dreslinski,
  Hsing~Min Chen, Robert Patti, Betina Hold, Chaitali Chakrabarti, Trevor
  Mudge, and David Blaauw.
\newblock Exploring dram organizations for energy-efficient and resilient
  exascale memories.
\newblock In \emph{Proceedings of the International Conference on High
  Performance Computing, Networking, Storage and Analysis}, pages 1--12, 2013.

\bibitem[Hassan et~al.(2019)Hassan, Patel, Kim, Yaglikci, Vijaykumar, Ghiasi,
  Ghose, and Mutlu]{hassan2019crow}
Hasan Hassan, Minesh Patel, Jeremie~S Kim, A~Giray Yaglikci, Nandita
  Vijaykumar, Nika~Mansouri Ghiasi, Saugata Ghose, and Onur Mutlu.
\newblock Crow: A low-cost substrate for improving dram performance, energy
  efficiency, and reliability.
\newblock In \emph{Proceedings of the 46th International Symposium on Computer
  Architecture}, pages 129--142, 2019.

\bibitem[Tomes and Altiparmak(2017)]{tomes2017comparative}
Erica Tomes and Nihat Altiparmak.
\newblock A comparative study of hdd and ssd raids’ impact on server energy
  consumption.
\newblock In \emph{2017 IEEE International Conference on Cluster Computing
  (CLUSTER)}, pages 625--626. IEEE, 2017.

\bibitem[Park et~al.(2011)Park, Kim, Urgaonkar, Lee, and
  Seo]{park2011comprehensive}
Seonyeong Park, Youngjae Kim, Bhuvan Urgaonkar, Joonwon Lee, and Euiseong Seo.
\newblock A comprehensive study of energy efficiency and performance of
  flash-based ssd.
\newblock \emph{Journal of Systems Architecture}, 57\penalty0 (4):\penalty0
  354--365, 2011.

\bibitem[Harris and Altiparmak(2020)]{harris2020ultra}
Bryan Harris and Nihat Altiparmak.
\newblock Ultra-low latency ssds' impact on overall energy efficiency.
\newblock \emph{USENIX HotStorage'20}, 2020.

\bibitem[Arunkumar et~al.(2019)Arunkumar, Bolotin, Nellans, and
  Wu]{arunkumar2019understanding}
Akhil Arunkumar, Evgeny Bolotin, David Nellans, and Carole-Jean Wu.
\newblock Understanding the future of energy efficiency in multi-module gpus.
\newblock In \emph{2019 IEEE International Symposium on High Performance
  Computer Architecture (HPCA)}, pages 519--532. IEEE, 2019.

\bibitem[Majumdar et~al.(2017)Majumdar, Piga, Paul, Greathouse, Huang, and
  Albonesi]{majumdar2017dynamic}
Abhinandan Majumdar, Leonardo Piga, Indrani Paul, Joseph~L Greathouse, Wei
  Huang, and David~H Albonesi.
\newblock Dynamic gpgpu power management using adaptive model predictive
  control.
\newblock In \emph{2017 IEEE International Symposium on High Performance
  Computer Architecture (HPCA)}, pages 613--624. IEEE, 2017.

\bibitem[Kandiah et~al.(2021)Kandiah, Peverelle, Khairy, Pan, Manjunath,
  Rogers, Aamodt, and Hardavellas]{kandiah2021accelwattch}
Vijay Kandiah, Scott Peverelle, Mahmoud Khairy, Junrui Pan, Amogh Manjunath,
  Timothy~G Rogers, Tor~M Aamodt, and Nikos Hardavellas.
\newblock Accelwattch: A power modeling framework for modern gpus.
\newblock In \emph{MICRO-54: 54th Annual IEEE/ACM International Symposium on
  Microarchitecture}, pages 738--753, 2021.

\bibitem[Leng et~al.(2013)Leng, Hetherington, ElTantawy, Gilani, Kim, Aamodt,
  and Reddi]{leng2013gpuwattch}
Jingwen Leng, Tayler Hetherington, Ahmed ElTantawy, Syed Gilani, Nam~Sung Kim,
  Tor~M Aamodt, and Vijay~Janapa Reddi.
\newblock Gpuwattch: Enabling energy optimizations in gpgpus.
\newblock \emph{ACM SIGARCH Computer Architecture News}, 41\penalty0
  (3):\penalty0 487--498, 2013.

\bibitem[Sethia and Mahlke(2014)]{sethia2014equalizer}
Ankit Sethia and Scott Mahlke.
\newblock Equalizer: Dynamic tuning of gpu resources for efficient execution.
\newblock In \emph{2014 47th Annual IEEE/ACM International Symposium on
  Microarchitecture}, pages 647--658. IEEE, 2014.

\bibitem[Khairy et~al.(2020)Khairy, Nikiforov, Nellans, and
  Rogers]{khairy2020locality}
Mahmoud Khairy, Vadim Nikiforov, David Nellans, and Timothy~G Rogers.
\newblock Locality-centric data and threadblock management for massive gpus.
\newblock In \emph{2020 53rd Annual IEEE/ACM International Symposium on
  Microarchitecture (MICRO)}, pages 1022--1036. IEEE, 2020.

\bibitem[Jahanshahi et~al.(2020)Jahanshahi, Sabzi, Lau, and
  Wong]{jahanshahi2020gpu}
Ali Jahanshahi, Hadi~Zamani Sabzi, Chester Lau, and Daniel Wong.
\newblock Gpu-nest: Characterizing energy efficiency of multi-gpu inference
  servers.
\newblock \emph{IEEE Computer Architecture Letters}, 19\penalty0 (2):\penalty0
  139--142, 2020.

\bibitem[Hu et~al.(2021)Hu, Sun, Yan, Wen, and Zhang]{hu2021characterization}
Qinghao Hu, Peng Sun, Shengen Yan, Yonggang Wen, and Tianwei Zhang.
\newblock Characterization and prediction of deep learning workloads in
  large-scale gpu datacenters.
\newblock In \emph{Proceedings of the International Conference for High
  Performance Computing, Networking, Storage and Analysis}, pages 1--15, 2021.

\bibitem[Patki et~al.(2019)Patki, Frye, Bhatia, Di~Natale, Glosli, Ingolfsson,
  and Rountree]{patki2019comparing}
Tapasya Patki, Zachary Frye, Harsh Bhatia, Francesco Di~Natale, James Glosli,
  Helgi Ingolfsson, and Barry Rountree.
\newblock Comparing gpu power and frequency capping: A case study with the
  mummi workflow.
\newblock In \emph{2019 IEEE/ACM Workflows in Support of Large-Scale Science
  (WORKS)}, pages 31--39. IEEE, 2019.

\bibitem[Grealey et~al.(2022)Grealey, Lannelongue, Saw, Marten, M{\'e}ric,
  Ruiz-Carmona, and Inouye]{grealey2022carbon}
Jason Grealey, Lo{\"\i}c Lannelongue, Woei-Yuh Saw, Jonathan Marten, Guillaume
  M{\'e}ric, Sergio Ruiz-Carmona, and Michael Inouye.
\newblock The carbon footprint of bioinformatics.
\newblock \emph{Molecular biology and evolution}, 39\penalty0 (3):\penalty0
  msac034, 2022.

\bibitem[Wu et~al.(2022)Wu, Raghavendra, Gupta, Acun, Ardalani, Maeng, Chang,
  Aga, Huang, Bai, et~al.]{wu2022sustainable}
Carole-Jean Wu, Ramya Raghavendra, Udit Gupta, Bilge Acun, Newsha Ardalani,
  Kiwan Maeng, Gloria Chang, Fiona Aga, Jinshi Huang, Charles Bai, et~al.
\newblock Sustainable ai: Environmental implications, challenges and
  opportunities.
\newblock \emph{Proceedings of Machine Learning and Systems}, 4:\penalty0
  795--813, 2022.

\bibitem[Patterson et~al.(2021)Patterson, Gonzalez, Le, Liang, Munguia,
  Rothchild, So, Texier, and Dean]{patterson2021carbon}
David Patterson, Joseph Gonzalez, Quoc Le, Chen Liang, Lluis-Miquel Munguia,
  Daniel Rothchild, David So, Maud Texier, and Jeff Dean.
\newblock Carbon emissions and large neural network training.
\newblock \emph{arXiv preprint arXiv:2104.10350}, 2021.

\bibitem[And{\'u}jar et~al.(2023)And{\'u}jar, Coll, Alonso, Mart{\'\i}nez,
  L{\'o}pez, S{\'a}nchez, and Alfaro]{andujar2023energy}
Francisco~J And{\'u}jar, Salvador Coll, Marina Alonso, Juan-Miguel
  Mart{\'\i}nez, Pedro L{\'o}pez, Jos{\'e}~L S{\'a}nchez, and Francisco~J
  Alfaro.
\newblock Energy efficient hpc network topologies with on/off links.
\newblock \emph{Future Generation Computer Systems}, 139:\penalty0 126--138,
  2023.

\bibitem[Georgakoudis et~al.(2019)Georgakoudis, Jain, Ono, Inoue, Miwa, and
  Bhatele]{georgakoudis2019evaluating}
Giorgis Georgakoudis, Nikhil Jain, Takatsugu Ono, Koji Inoue, Shinobu Miwa, and
  Abhinav Bhatele.
\newblock Evaluating the impact of energy efficient networks on hpc workloads.
\newblock In \emph{2019 IEEE 26th International Conference on High Performance
  Computing, Data, and Analytics (HiPC)}, pages 301--310. IEEE, 2019.

\bibitem[Dietrich and Chien(2022)]{dietrich2022navigating}
Mark Dietrich and Andrew Chien.
\newblock Navigating dennard, carbon and moore: Scenarios for the future of nsf
  advanced computational infrastructure.
\newblock In \emph{Practice and Experience in Advanced Research Computing},
  pages 1--6. 2022.

\end{thebibliography}
